\documentclass[11pt,a4paper]{article}
\pdfoutput=1
\usepackage{jheppub} 
\usepackage{amsfonts,amssymb,latexsym,amsmath,mathrsfs,amsbsy}
\usepackage{bm}
\usepackage{slashed,bbold}
\usepackage[hhmmss]{datetime}
\usepackage{dsfont}

\newcommand{\mcG}{\mathcal G}

\newcommand{\mbJ}{\bm{J}}
\newcommand{\mcB}{\mathcal B}
\newcommand{\mbB}{\bm{B}}
\newcommand{\mbF}{\bm{F}}
\newcommand{\mbA}{\bm{A}}
\newcommand{\mbV}{\bm{V}}
\newcommand{\bsA}{\boldsymbol{A}}
\newcommand{\bsF}{\boldsymbol{F}}
\newcommand{\mcA}{{\mathcal A}}

\newcommand{\mcJ}{{\mathcal J}}

\newcommand{\mcV}{{\mathcal V}}

\newcommand{\mcF}{{\mathcal F}}
\newcommand{\mcE}{{\mathcal E}}

\newcommand{\x}{{\bf x}}
\newcommand{\y}{{\bf y}}
\newcommand{\Tr}{{\textrm {Tr}}}

\usepackage[latin1]{inputenc}

\title{Anomalous Currents and Constitutive Relations of a Chiral Hadronic Superfluid} 

\author[a]{Juan L. Ma\~nes,}
\author[b,c]{Eugenio Meg\'\i as,}
\author[b]{Manuel Valle,}
\author[d]{Miguel \'A. V\'azquez-Mozo}

\affiliation[a]{Departamento de F\'\i sica de la Materia Condensada, 
Universidad del Pa\'is Vasco UPV/EHU, \\
Apartado 644,  48080 Bilbao, Spain}
\affiliation[b]{Departamento de F\'\i sica Te\'orica, 
Universidad del Pa\'is Vasco UPV/EHU, \\
Apartado 644,  48080 Bilbao, Spain}
\affiliation[c]{Departamento de F\'{\i}sica At\'omica, Molecular y Nuclear and Instituto Carlos I de F\'{\i}sica Te\'orica y Computacional, Universidad de Granada, Avenida de Fuente Nueva s/n,  18071 Granada, Spain}
\affiliation[d]{Departamento de F\'\i sica Fundamental, Universidad de Salamanca, \\
Plaza de la Merced s/n, 37008 Salamanca, Spain}

\emailAdd{wmpmapaj@lg.ehu.es}
\emailAdd{emegias@ugr.es}
\emailAdd{manuel.valle@ehu.es}
\emailAdd{Miguel.Vazquez-Mozo@cern.ch}

\abstract{
The anomalous currents of two-flavor chiral nuclear matter in the presence of chiral imbalance are computed, 
using recently developed methods exploiting generalized transgression, which facilitates the 
evaluation of both the equilibrium partition function and the covariant currents. The constitutive
relations for both the broken and unbroken phase of the theory are studied and the out-of-equilibrium
nondissipative transport coefficients determined. In the superfluid phase, 
the vector covariant currents exhibit nondissipative chiral electric, magnetic, and vortical effects, the latter governed by chiral imbalance. 
}

\arxivnumber{}

\begin{document}

\maketitle


\flushbottom


\section{Introduction}
\label{sec:intro}

Quantum chiral anomalies are associated with very robust mathematical properties of gauge fields
beyond perturbation theory
\cite{Zumino:1983ew,AlvarezGaume:1985ex,Bertlmann:1996xk,Fujikawa:2004cx,Harvey:2005it,Bilal:2008qx}.
A practical consequence of this is that the anomalous contribution to the quantum effective action can be 
obtained from very general considerations using differential geometry. Although nonlocal in (Euclidean) dimension $D=2n-2$, 
the effective action can be written as a local functional in $D+1=2n-1$ dimensions, using the Chern-Simons form 
associated with the appropriate anomaly polynomial. 
For systems exhibiting spontaneous symmetry breaking, the mathematical structure of the anomaly 
determines the Wess-Zumino-Witten (WZW) effective Lagrangian describing the low-energy interaction of Nambu-Goldstone (NG) bosons with 
gauge fields. 

This state of affairs turns out to be specially useful in the study of 
anomalous hydrodynamic transport \cite{Son:2009tf,Landsteiner:2016led}, where fluids are coupled to classical external sources
through currents affected by 't Hooft anomalies. 
These novel class of transport phenomena include the chiral magnetic \cite{Fukushima:2008xe,Sadofyev:2010pr}
and electric \cite{Neiman:2011mj}
effects, as well as
the chiral vortical effect \cite{Kirilin:2012mw} (see also \cite{Fukushima:2012vr,Zakharov:2012vv} for dedicated reviews). 
They are of physical relevance in various physical setups, including heavy ion collisions~\cite{Kharzeev:2010gr,Hongo:2013cqa,Kharzeev:2015znc,Huang:2015oca} as well as 
astrophysics and cosmology~\cite{Kaminski:2014jda,Sigl:2015xva,Yamamoto:2015gzz,Masada:2018swb,Abbaslu:2019yiy}.

The equilibrium partition functions at finite temperature for these systems can be computed
from the anomalous effective action 
functional by implementing dimensional reduction on the Euclidean time cycle \cite{Jensen:2013kka}. The anomaly 
is then carried by the local part of the partition function while the covariant currents, giving the response of the
system to the physical sources, can be obtained
from its nonlocal piece. Transport coefficients are then read off the expressions of these currents. The application of these
techniques to the study of anomalous hydrodynamics has been developed in a 
number of works~\cite{Jensen:2013kka,Jensen:2012kj,Jensen:2013rga,Haehl:2013hoa,Monteiro:2014wsa,Jain:2015jla,Banerjee:2015hra,Glorioso:2017lcn}.

In a previous paper \cite{Manes:2018llx}, we used differential geometry to obtain explicit operative expressions allowing the
computation of the equilibrium partition function of fluids in the presence of quantum ('t Hooft) anomalies, 
as well as the construction of the 
various currents associated with the corresponding global symmetries. Although applicable to generic theories, the strategy presented there
turned out to be particularly powerful for systems with spontaneous symmetry breaking, where the covariant anomalous 
currents can be obtained without going through the study of the WZW effective action \cite{Fukushima:2012fg,Brauner:2018zwr}, which
generically has a rather cumbersome expression. 
In fact, these currents can be readily computed by an appropriate transformation of the Bardeen-Zumino (BZ) 
terms of the unbroken theory using the NG boson matrix.  
This method has been recently implemented in the study of multi-Weyl semimetals~\cite{Dantas:2019rgp}.

In this work we show how the results of Ref. \cite{Manes:2018llx} are applied to the analysis of chiral nuclear matter (super)fluids 
in the presence of chiral, isospin, and baryon number imbalance. More specifically, we study a two-flavor hadronic fluid with unbroken chiral symmetry,
as well as its superfluid phase when this symmetry is spontaneously broken.  
In the latter case, the equilibrium partition function is computed from the WZW functional 
by implementing dimensional reduction. The standard computation of the consistent currents would proceed by taking functional derivatives
of this complicated action with respect to classical external sources. 
The corresponding covariant (vector and axial-vector) currents would then be obtained by adding the appropriate BZ terms.
Instead of doing this, here we compute the covariant currents directly from the BZ polynomials using the explicit formulae found in~\cite{Manes:2018llx},
which notably simplifies the calculation.

By expressing the currents in terms of the different Lorentz tensor structures built from the NG bosons, together with the
magnetic field and the vorticity, we identify  
the different transport phenomena present in the theory and compute the corresponding out-of-equilibrium transport coefficients. 
Our results show the existence of the chiral electric effect 
first found in \cite{Neiman:2011mj} and confirm its nondissipative nature. 
In addition to this, we also find that all dependence of the electromagnetic, baryonic and isospin currents on the vorticity
comes weighted by the chemical potential $\mu_{5}$ and therefore disappears in the absence of chiral imbalance $\mu_{5}\rightarrow 0$. 
Despite being erased from the currents, vorticity dependent terms survive in the equilibrium number densities in this limit. 
Explicit expressions of the out-of-equilibrium transport coefficients in the unbroken phase of the theory are also provided.

The article is organized as follows. In Section \ref{sec:anomaly} we briefly review the main results of Ref. \cite{Manes:2018llx}
which will be later used in the analysis of hadronic (super)fluids. After this, we study in Section
\ref{sec:2fQCD_unbroken} a two-flavor hadronic fluid without chiral symmetry breaking and compute its equilibrium partition function, as
well as the relevant currents and the energy-momentum tensor to leading and first order in the derivative expansion. Section \ref{sec:pion_part_funct} will be devoted
to the study of this system in the phase in which chiral symmetry is spontaneously broken, 
computing the corresponding pion partition function up to first order in derivatives. In Section \ref{sec:currents}, the covariant currents and constitutive relations of this hadron superfluid 
will be computed, and the emergence of various chiral effects discussed. Finally, we summarize our conclusions in Section \ref{sec:discussion}.

\section{Equilibrium partition functions from differential geometry}
\label{sec:anomaly}

For the reader's convenience and to make our presentation self-contained, in this section 
we briefly summarize the formalism and main results of Ref. \cite{Manes:2018llx}. Key expressions 
will be given in components for their use later on in the paper. 

We study a theory of chiral fermions coupled to external gauge fields focusing for the time being on the simplest case of a single gauge
potential one-form and its field strength two-form, both taking values on the algebra of the gauge group $\mcG$
\begin{align}
\mcA&=-i\mcA_{\mu}dx^{\mu}, \nonumber \\[0.2cm]
\mcF&\equiv d\mcA+\mcA^{2}=-{i\over 2}\mcF_{\mu\nu}dx^{\mu}dx^{\nu}.
\label{eq:defs_AF_cal1&2}
\end{align} 
The main ingredient in the construction of the anomalous effective action in $D=2n-2$ dimensions, and thus of the equilibrium partition function, 
is the anomaly polynomial, 
a $2n$-form built in terms of $\mcF$. If the external gauge field $\mcA$ is 
coupled to a right-handed fermion, this is given by
\begin{align}
\mathcal{P}(\mcF)&\equiv 2\pi({\rm ch\,}\mcF)_{2n}=c_{n}{\rm Tr\,}\mcF^{n} \hspace*{1cm} \mbox{with}
\hspace*{1cm} c_{n}={1\over n!}{i^{n}\over (2\pi)^{2n-1}},
\end{align}
where ${\rm ch\,}\mcF$ is the Chern character (see \cite{Nakahara:2003nw} for definitions and conventions) and its subscript indicates that we only retain the form of degree $2n$. The global
normalization is dictated by the Atiyah-Singer index theorem relating the anomaly with the topological properties of the gauge bundle.
The anomalous part of the effective action $\Gamma[\mcA]_{\rm CS}$ can be constructed then
in terms of the Chern-Simons form $\omega^{0}_{2n-1}(\mcA,\mcF)$, defined by
\begin{align}
\mathcal{P}(\mcF)=d\omega^{0}_{2n-1}(\mcA,\mcF) \hspace*{1cm} \Longrightarrow \hspace*{1cm}
\Gamma[\mcA]_{\rm CS}=c_{n}\int\limits_{\mathcal{M}_{2n-1}}\omega^{0}_{2n-1}(\mcA,\mcF).
\label{eq:CS_action_general}
\end{align}
In the case of a left-handed fermion, the corresponding expression for the effective action carries an additional global minus sign. 
For theories with various chiral fermions, the result is
obtained by adding the contributions of the different species weighted by the corresponding signs ($+$ for right- and $-$ for left-handed
fermions).

To compute the equilibrium partition function, we define the theory on a 
generic static background~\cite{Banerjee:2012iz} whose metric, upon choosing appropriate coordinates, can be
brought to the form
\begin{align}
ds^{2}&\equiv G_{\mu\nu}(\x)dx^{\mu}dx^{\nu}=-e^{2\sigma(\x)}\Big[dt+a_{i}(\x)dx^{i}\Big]^{2}+g_{ij}(\x)dx^{i}dx^{j}.
\label{eq:static_line_element}
\end{align}
Written like this, the metric has manifest timelike Killing vector given by $\partial_{t}$. We also take all other tensor fields in the theory 
to have vanishing Lie derivative with respect to this Killing
field, which in the coordinates chosen simply means that they are time-independent.

The form of the line element \eqref{eq:static_line_element} is preserved by 
Kaluza-Klein (KK) transformations, defined by position-dependent shifts generated by the vector field $\zeta(\x)=\phi(\x)\partial_{t}$
together 
with an appropriate redefinition of the metric function $a_{i}(\x)$, namely
\begin{align}
(t,\x)&\longrightarrow (t+\phi,\x ), \nonumber \\[0.2cm]
a_{i}&\longrightarrow a_{i}-\partial_{i}\phi.
\label{eq:KK_transf}
\end{align}
The diffeomorphism generated by $\zeta(\x)$ induces the following change on the components of gauge field one-forms defined on this spacetime
\begin{align}
\mcA_{0}& \longrightarrow \mcA_{0}, \nonumber \\[0.2cm]
\mcA_{i}& \longrightarrow \mcA_{i}-\mcA_{0}\partial_{i}\phi.
\label{eq:AKKgauge_trans}
\end{align}
Associated with this metric, we can introduce the properly normalized four-velocity
of comoving observers in the static metric \eqref{eq:static_line_element}
\begin{align}
u^{\mu}=e^{-\sigma}\big(1,\mathbf{0}\big) \hspace*{1cm} \Longrightarrow \hspace*{1cm}
u_{\mu}=-e^{\sigma}\big(1,a_{i}\big).
\label{eq:four_velocity}
\end{align}
This defines a congruence of timelike geodesics foliating the spacetime by spatial hypersurfaces normal
to $u^{\mu}$, whose induced metric is given by
\begin{align}
h_{\mu\nu}=G_{\mu\nu}+u_{\mu}u_{\nu} \hspace*{1cm} \Longrightarrow \hspace*{1cm}
h_{ij}=g_{ij},
\end{align}
with $h_{00}=h_{0i}=0$. 

Tensor fields in this geometry can be decomposed into KK-invariant combinations by splitting them into longitudinal and transverse components
with respect to the vector $u^{\mu}$. To do this, we introduce the projector
\begin{align}
P^{\mu}_{\,\,\,\nu}\equiv h^{\mu}_{\,\,\,\nu}=\delta^{\mu}_{\,\,\,\nu}+u^{\mu}u_{\nu}
=\left(
\begin{array}{cr}
0 & -a_{i} \\
0 & \delta^{i}_{\,j}
\end{array}
\right),
\label{eq:projector_def}
\end{align}
and contract indices with the identity written in the form $\delta^{\mu}_{\,\,\,\nu}=-u^{\mu}u_{\nu}+P^{\mu}_{\,\,\,\nu}$. In the case of
a gauge field, the result is
\begin{align}
\mcA_{\mu}=-\big(\mcA_{\nu}u^{\nu}\big)u_{\mu}+\mcA_{\nu}P^{\nu}_{\,\,\,\mu}.
\end{align}
Scalar quantities are, by definition, invariant under KK transformations and so is the combination appearing in the first term
\begin{align}
\mcA_{\mu}u^{\mu}=e^{-\sigma}\mcA_{0}\equiv e^{-\sigma}A_{0}.
\end{align}
For the transverse components, using the explicit form of the projector, we find 
\begin{align}
\mcA_{\nu}P^{\nu}_{\,\,\,\mu}=\big(0,\mcA_{i}-\mcA_{0}a_{i}\big)\equiv \big(0,A_{i}\big).
\end{align}
This pulled-back one-form, having a vanishing time component, is automatically KK-invariant as can be checked from Eq.~\eqref{eq:AKKgauge_trans}.
Thus, gauge field one-form in Eq. \eqref{eq:defs_AF_cal1&2} can be written in terms of KK-invariant components defined by
\begin{align}
A_{0}&\equiv e^{\sigma} u^{\mu}\mcA_{\mu}=\mcA_{0}, \nonumber \\[0.2cm]
A_{i}&\equiv \mcA_{\mu}P^{\mu}_{\,\,\,i}=\mcA_{i}-\mcA_{0}a_{i}.
\label{eq:def_ttgaugefield_gen}
\end{align}
From here we retrieve the decomposition used in \cite{Manes:2018llx}
\begin{align}
\mcA&=-i\mcA_{\nu}\Big(-u^{\nu}u_{\mu}+P^{\nu}_{\,\,\,\mu}\Big)dx^{\mu}
\equiv -i\mcA_{0}\theta(\x)+\mbA=-iA_{0}\theta-iA_{i}dx^{i},
\label{eq:KK_inv_gaugefields}
\end{align}
where we have defined the one-form $\theta\equiv -e^{-\sigma}u_{\mu}dx^{\mu}$. Together with $dx^{i}$, it makes up a KK-invariant cotangent basis\footnote{The KK-invariant
gauge one-form $ A_{i}dx^{i}$ corresponds to the hatted connection used in Ref. \cite{Jensen:2013kka}.}.
The relations \eqref{eq:def_ttgaugefield_gen} can be inverted to write
\begin{align}
\mcA_{\mu}=-e^{-\sigma}A_{0}u_{\mu}+A_{i}\delta^{i}_{\mu}.
\label{eq:general_decomp_Afield}
\end{align}
Similar decompositions can be written also for a vector field $\mathcal{W}^{\mu}\partial_{\mu}$ in terms of KK-invariant
components $(W^{0},W^{i})$, by expanding it in the KK-invariant tangent basis 
$\{u^{\mu}\partial_{\mu},P^{\mu}_{\,\,\,i}\partial_{\mu}\}=\{e^{\sigma}\partial_{0},\partial_{i}-a_{i}\partial_{0}\}$
\begin{align}
\mathcal{W}^{\mu}\partial_{\mu}&=-\big(\mathcal{W}^{\nu}u_{\nu}\big)u^{\mu}\partial_{\mu}+P^{\mu}_{\,\,\,i}\mathcal{W}^{i}\partial_{\mu}
\nonumber \\[0.2cm]
&=e^{\sigma}W^{0}\partial_{0}+W^{i}\big(\partial_{i}-a_{i}\partial_{0}\big)
=\big(e^{\sigma}W^{0}-W^{i}a_{i}\big)\partial_{0}+W^{i}\partial_{i},
\label{eq:vector_KK-inv_decomp_pre}
\end{align}
so we have
\begin{align}
\mathcal{W}^{\mu}=\big(e^{\sigma}W^{0}-W^{i}a_{i}\big)u^{\mu}+W^{i}\delta^{\mu}_{i}.
\label{eq:vector_KK-inv_decomp}
\end{align}

The notation used in this paper is a slight modification of the one introduced in Ref.~\cite{Manes:2018llx}. 
We reserve calligraphic script $\mcA$, $\mcF$,... 
to denote the {\em KK-variant} components. 
Roman script $A,F,\ldots$ indicates the corresponding {\em KK-invariant} fields.  
From Eq. \eqref{eq:AKKgauge_trans}, we notice that the time component of the gauge field $\mcA_{0}$ is KK invariant, so
we have $\mcA_{0}=A_{0}$. However, unlike in Ref.~\cite{Manes:2018llx}, here we only use {\em Hermitian} components for the
diverse adjoint fields, as can be seen from the presence of the $-i$ factor in Eq. \eqref{eq:defs_AF_cal1&2}.

To construct the equilibrium partition function at temperature $T_{0}\equiv \beta^{-1}$, we use the imaginary time formalism setting in 
\eqref{eq:CS_action_general} a
manifold $\mathcal{M}_{2n-1}$ with topology $S^{1}\times D_{2n-2}$, where the second factor is a $(2n-2)$-dimensional 
ball and the thermal cycle $S^{1}$ has length $\beta$. 
Since all fields are time-independent, the integration over $dx^{0}$ is
trivially implemented by replacing
\begin{align}
i\hspace*{-0.4cm}\int\limits_{S^{1}\times D_{2n-2}}\hspace*{-0.4cm}dt \longrightarrow {1\over T_{0}}\int\limits_{D_{2n-2}}.
\label{eq:dim_red_gen}
\end{align}
Using this prescription, one finds that the equilibrium partition function has the following structure
\begin{align}
i\Gamma[\mcA_{0},\bsA]_{\rm CS}=W[\mcA_{0},\bsA]_{\rm inv}+W[\mcA_{0},\bsA]_{\rm anom},
\label{eq:inv+anom_decomp}
\end{align}
where the first piece is gauge invariant and the anomaly is carried by the second term.
This decomposition was found in \cite{Jensen:2013kka}, while in \cite{Manes:2018llx} we provided closed expressions for both terms
using generalized transgression \cite{Manes:1985df}. 
The first, gauge invariant piece can be written as the following integral over the bulk manifold $D_{2n-2}$
\begin{align}
W[\mcA_{0},\bsA]_{\rm inv}&=-{inc_{n}\over T_{0}}\int\limits_{D_{2n-2}}\int_{0}^{1}dt\,{\rm Tr\,}\Big[\mcA_{0}\big(\bsF-it\mcA_{0}da\big)^{n-1}\Big],
\label{eq:prescription_inv}
\end{align}
where we have introduced the KK gauge field one-form $a\equiv a_{i}dx^{i}$. Being a higher-dimensional integral,
this term is nonlocal in physical space. The anomalous part of the partition function, on the other hand, takes the form
\begin{align}
W[\mcA_{0},\bsA]_{\rm anom}=
\left.-{ic_{n}\over T_{0}}\int\limits_{S^{2n-3}}\int_{0}^{1}dt\,\mcA_{0}{\delta\over\delta \mcF}\omega^{0}_{2n-1}(\mcA,\mcF)
\right|_{\substack{\hspace*{-1cm}\mcA\to\mbA \\[0.1cm]\mcF\to\mbF-it\mcA_0da}},
\label{eq:prescription_anom}
\end{align}
where $S^{2n-3}\equiv \partial D_{2n-2}$ is the physical space. This contribution to the effective action 
not only carries the anomaly, but is actually local. 

It should be stressed at this point that, despite its gauge invariance, 
the nonlocal piece in \eqref{eq:inv+anom_decomp} is crucial for the computation of the covariant currents and the
anomalous contribution to the energy-momentum tensor. Indeed, covariant and consistent currents are
then derived from the variation of the invariant and anomalous pieces of the partition function under  
$\delta_{B}\mcA=\mcB_{0}\theta+\mbB$ 
\begin{align}
\delta_{B}W[\mcA_{0},\mbA]_{\rm inv}&=\int\limits_{S^{2n-3}}{\rm Tr\,}\Big(-i\mcB_{0}\mcJ_{0,{\rm cov}}+\mbB\mbJ_{\rm cov}\Big)
+\mbox{bulk terms}, \nonumber \\[0.2cm]
\delta_{B}W[\mcA_{0},\mbA]_{\rm anom}&=\int\limits_{S^{2n-3}}{\rm Tr\,}\Big(-i\mcB_{0}\mcJ_{0,{\rm cons}}+\mbB\mbJ_{\rm cons}\Big).
\end{align}

In fact, the local character of the anomalous part of the effective action in \eqref{eq:inv+anom_decomp} can be derived on more general grounds.
To fix ideas, let us consider the simplest case of a left- and a right-handed fermion coupled to external U(1) vector 
$\mathcal{V}_{\mu}=(\mcV_{0},\boldsymbol{\mcV})$ and 
axial-vector $\mcA_{\mu}=(\mcA_{0},\boldsymbol{\mcA})$ 
external gauge fields in flat four-dimensional spacetime. 
The nonlocal anomalous effective action for these fields can be written as~\cite{Giannotti:2008cv}
\begin{align}
S[\mcV,\mcA]_{\rm eff}=-{e^{2}\over 16\pi^{2}}\int d^{4}x\int d^{4}y\,\Big(\epsilon^{\mu\nu\alpha\beta}\mcV_{\mu\nu}\mcV_{\alpha\beta}\Big)_{x}
\left({1\over \Box}\right)_{xy}\Big(\partial_{\sigma}\mcA^{\sigma}\Big)_{y},
\end{align}
where $\mcV_{\mu\nu}=\partial_{\mu}\mcV_{\nu}-\partial_{\nu}\mcV_{\mu}$ is the field strength associated with the Abelian vector gauge field. 
In the static case where all fields are time-independent, we can implement dimensional reduction and the effective action simplifies to
\begin{align}
S[\mcV,\mcA]_{\rm eff}&={ie^{2}\over 4\pi^{2}T_{0}}\int d^{3}\x\int d^{3}\y\,\Big(\epsilon^{ijk}\mcV_{ij}\mcV_{0k}\Big)_{\x}
\left({1\over \boldsymbol{\nabla}^{2}}\right)_{\x\y}
\Big(\boldsymbol{\nabla}\cdot\boldsymbol{\mcA}\Big)_{\y} \nonumber \\[0.2cm]
&=-{ie^{2}\over 4\pi^{2}T_{0}}\int d^{3}\x\int d^{3}\y\,\Big(\boldsymbol{\nabla}\mcV_{0}\cdot\boldsymbol{\mathcal{B}}\Big)_{\x}
\left({1\over \boldsymbol{\nabla}^{2}}\right)_{\x\y}
\Big(\boldsymbol{\nabla}\cdot\boldsymbol{\mcA}\Big)_{\y},
\end{align}
where we have identified the electric and magnetic fields associated to the vector gauge connection,
$\boldsymbol{\mathcal{E}}=-\boldsymbol{\nabla}\mcV_{0}$ and $\boldsymbol{\mathcal{B}}=\boldsymbol{\nabla}\times\boldsymbol{\mcV}$. 
Taking into account that $\boldsymbol{\nabla}\cdot\boldsymbol{\mathcal{B}}=0$ and
integrating by parts twice, we find
\begin{align}
S[\mcV,\mcA]_{\rm eff}&=-{ie^{2}\over 4\pi^{2}T_{0}}\int d^{3}\x\int d^{3}\y\,\Big(\mcV_{0}\,\boldsymbol{\mathcal{B}}\Big)_{\x}\left[
\boldsymbol{\nabla}_{\y}\cdot\boldsymbol{\nabla}_{\x}
\left({1\over \boldsymbol{\nabla}^{2}}\right)_{\x\y}\right]\boldsymbol{\mcA}_{\y} \nonumber \\[0.2cm]
&=-{ie^{2}\over 4\pi^{2}T_{0}}\int d^{3}\x\,\Big(\mcV_{0}\,\boldsymbol{\mathcal{B}}\cdot\boldsymbol{\mcA}\Big)_{\x},
\end{align}
which is indeed a local functional in three dimensions. This expression is invariant under vector gauge transformations (notice that $\mcV_{0}$ 
does not change when the gauge parameter is time-independent), whereas its variation 
under axial-vector gauge transformations $\boldsymbol{\mcA}\rightarrow
\boldsymbol{\mcA}+\boldsymbol{\nabla}\eta$ reproduces the axial anomaly
\begin{align}
\delta_{\eta}S[\mcV,\mcA]_{\rm eff}=-{ie^{2}\over 4\pi^{2}T_{0}}\int d^{3}\x\,\eta\,\boldsymbol{\mathcal{E}}\cdot\boldsymbol{\mathcal{B}}.
\end{align}

In Ref. \cite{Manes:2018llx} a general prescription was given allowing the construction of the equilibrium partition function of generic theories,
as well their consistent and covariant currents. Concerning the energy-momentum tensor,  
the Chern-Simons effective action \eqref{eq:CS_action_general} is a topological invariant and therefore independent of the metric
defined on the manifold $\mathcal{M}_{2n-1}$. Naively, it would seem then that the anomaly does not induce any terms in the energy-momentum tensor. 
However, this is not so. As a result of the dimensional reduction \eqref{eq:dim_red_gen} based on the KK-invariant decomposition \eqref{eq:KK_inv_gaugefields}, 
the partition function acquires an explicit dependence on the metric function~$a_{i}(\x)$, which brings about a nonzero anomalous
contribution to the $T_{0}^{\,\,\,i}$ component of the energy-momentum tensor.

\paragraph{The Bardeen anomaly.}
From now on we focus our analysis on the case of the Bardeen anomaly in $D=4$ ($n=3$), where the corresponding
equilibrium partition function depends on non-Abelian vector $\mcV=-i\mcV_{\mu}dx^{\mu}$ 
and axial-vector $\mcA=-i\mcA_{\mu}dx^{\mu}$ 
external gauge fields. 
The appropriate Chern-Simons form
preserving the invariance under vector gauge transformations takes the form~(see \cite{Manes:2018llx} for details)
\begin{align}
\widetilde{\omega}^{0}_{5}(\mcA,\mcF_{V},\mcF_{A})=6{\rm Tr\,}\left(\mcA\mcF_{V}^{2}+{1\over 3}\mcA\mcF_{A}^{2}
-{4\over 3}\mcA^{3}\mcF_{V}+{8\over 15}\mcA^{5}\right),
\label{eqw:omega05-tilde}
\end{align}
where the field strengths are given in components by
\begin{align}
\mcF_{V}&=-{i\over 2}\mcV_{\mu\nu}dx^{\mu}dx^{\nu}
\equiv -{i\over 2}\Big(\partial_{\mu}\mcV_{\nu}-\partial_{\nu}\mcV_{\mu}
-i[\mcV_{\mu},\mcV_{\nu}]-i[\mcA_{\mu},\mcA_{\nu}]\Big)dx^{\mu}dx^{\nu}, \nonumber \\[0.2cm]
\mcF_{A}&=-{i\over 2}\mcA_{\mu\nu}dx^{\mu}dx^{\nu}
\equiv -{i\over 2}\Big(\partial_{\mu}\mcA_{\nu}-\partial_{\nu}\mcA_{\mu}-i[\mcV_{\mu},\mcA_{\nu}]-i[\mcA_{\mu},\mcV_{\nu}]\Big)dx^{\mu}dx^{\nu}.
\label{eq:FV_FA_nonKK_defs}
\end{align}
After implementing dimensional reduction and writing everything in terms of the corresponding KK-invariant fields 
$(\mcV_{0},V_{i})$ and $(\mcA_{0},A_{i})$
[cf. \eqref{eq:KK_inv_gaugefields}], we find that 
the invariant and anomalous parts of the anomalous functional can be computed
by the likes of Eqs. \eqref{eq:prescription_inv} and \eqref{eq:prescription_anom} (details and the explicit expressions can be found in Ref. \cite{Manes:2018llx}). 
The expectation values of the two-forms dual to the 
spatial vector and axial-vector covariant
currents 
can be obtained by functional differentiation from the
invariant nonlocal part of the partition function according to \cite{Jensen:2013kka,Manes:2018llx}
\begin{align}
\langle\boldsymbol{J}_{V}\rangle_{\rm cov}&\equiv \langle\star\boldsymbol{j}_{V}\rangle_{\rm cov} 
=T_{0}{\delta\over\delta \boldsymbol{F}_{V}}W[\mcV_{0},\mcA_{0},\boldsymbol{F}_{V},\boldsymbol{F}_{A}]_{\rm inv},
\nonumber \\[0.2cm]
\langle\boldsymbol{J}_{A}\rangle_{\rm cov}&\equiv \langle\star\boldsymbol{j}_{A}\rangle_{\rm cov}
=T_{0}{\delta\over\delta \boldsymbol{F}_{A}}W[\mcV_{0},\mcA_{0},\boldsymbol{F}_{V},\boldsymbol{F}_{A}]_{\rm inv},
\end{align}
where the star represents the Hodge dual on the spatial $S^{3}$. An explicit expression of these currents has been given in 
Ref.~\cite{Manes:2018llx}. Writing
\begin{align}
\boldsymbol{j}=-i\mcJ_{k}dx^{k},
\label{eq:current_ifactor}
\end{align}
and expressing the KK-invariant vector and axial-vector gauge forms in terms of their components
\begin{align}
\begin{array}{ll}
\displaystyle \boldsymbol{V}&\displaystyle=-iV_{i}dx^{i}, \\[0.4cm]
\displaystyle\boldsymbol{A}&\displaystyle=-iA_{i}dx^{i},
\end{array}
\hspace*{1cm} \Longrightarrow \hspace*{1cm}
\begin{array}{ll}
\displaystyle \boldsymbol{F}_{V}&=\displaystyle-{i\over 2}V_{ij}dx^{i}dx^{j}, \\[0.4cm]
\displaystyle\boldsymbol{F}_{A}&=\displaystyle-{i\over 2}A_{ij}dx^{i}dx^{j},
\end{array}
\label{eq:forms_in_components}
\end{align}
we arrive at the expressions
\begin{align}
\langle\mcJ^{i}_{aV}\rangle_{\rm cov}&={N_{c}\over 8\pi^{2}}e^{-\sigma}\epsilon^{ik\ell}{\rm Tr\,}\Big\{t_{a}
\Big[V_{0}A_{k\ell}+A_{k\ell}V_{0}
+A_{0}V_{k\ell}
+V_{k\ell}A_{0}+2(V_{0}A_{0}+A_{0}V_{0})\partial_{k}a_{\ell}\Big]\Big\},  \nonumber \\[0.2cm]
\langle\mcJ^{i}_{aA}\rangle_{\rm cov}&={N_{c}\over 8\pi^{2}}e^{-\sigma}\epsilon^{ik\ell}{\rm Tr\,}\Big\{t_{a}
\Big[V_{0}V_{k\ell}+V_{k\ell}V_{0}
+A_{0}A_{k\ell}
+A_{k\ell}A_{0}+2(V_{0}^{2}+A_{0}^{2})\partial_{k}a_{\ell}\Big]\Big\},
\label{eq:currentsJ_vJ_a}
\end{align}
where, for later use, we have assumed that chiral fermions coupling to the external 
fields come in $N_{c}$ different colors.
To keep the notation homogeneous, we used 
$\mcV_{0}= V_{0}$ and $\mcA_{0}= A_{0}$. 
In addition to this, the covariant time components are zero for both the vector and axial-vector covariant 
currents~\cite{Jensen:2013kka,Manes:2018llx}
\begin{align}
\langle\mcJ_{a0V}\rangle_{\rm cov}&=\langle\mcJ_{a0A}\rangle_{\rm cov}=0.
\end{align}
Raising the index using the static metric \eqref{eq:static_line_element}, we find
\begin{align}
\langle\mcJ_{aV}^{0}\rangle_{\rm cov}=-a_{i}\langle\mcJ_{aV}^{i}\rangle_{\rm cov}, \nonumber \\[0.2cm]
\langle\mcJ_{aA}^{0}\rangle_{\rm cov}=-a_{i}\langle\mcJ_{aA}^{i}\rangle_{\rm cov},
\label{eq:currentsJ_vJ_a_0comp}
\end{align}
where the right-hand side of these expressions can be read off from Eq. \eqref{eq:currentsJ_vJ_a} and spatial indices are raised
and lowered
using the three-dimensional transverse metric $g_{ij}$. The flavor components of the current 
are defined by
\begin{align}
\mcJ^{\mu}_{a}\equiv {\rm Tr\,}\Big(t_{a}\mcJ^{\mu}\Big),
\end{align}
with $t_{a}$ the gauge group generators.
Identities \eqref{eq:currentsJ_vJ_a_0comp} can be written in a Lorentz-covariant way by stating that 
the covariant current is transverse to the four-velocity, namely
\begin{align}
u_{\mu}\langle\mcJ_{aV}^{\mu}\rangle_{\rm cov}=u_{\mu}\langle\mcJ_{aA}^{\mu}\rangle_{\rm cov}=0.
\end{align}

As for the anomaly-induced component of the energy-momentum tensor, its dual
two-form can be also obtained by taking functional derivatives of the invariant part of the effective action
\cite{Manes:2018llx}
\begin{align}
\boldsymbol{T}(\mcV_{0},\mbV,\mcA_{0},\mbA)
&=T_{0}\left[{\delta\over \delta (da)}+i\mcV_{0}{\delta\over\delta\boldsymbol{F}_{V}}+i\mcA_{0}{\delta\over \delta
\boldsymbol{F}_{A}}\right]W[\mcV_{0},\mcA_{0},\boldsymbol{F}_{V},\boldsymbol{F}_{A}]_{\rm inv}.
\end{align}
Taking the Hodge dual and writing the equation in components, we find the explicit expression
\begin{align}
T_{0}^{\,\,\,i}(V_0,A_0,V_{i},A_{i})&=-{N_{c}\over 8\pi^{2}}
e^{-\sigma}\epsilon^{ijk}{\rm Tr\,}\bigg[V_{jk}( V_{0}A_{0}+A_{0}V_{0})
+A_{jk}(V_{0}^{2}+A_{0}^{2})
\nonumber\\[0.2cm]
&\left.+{2\over 3}\partial_{j}a_{k}(A_{0}^{3}+3A_{0}V_{0}^{2})\right].
\label{eq:EM_tensor_Bardeen}
\end{align}

\paragraph{Nambu-Goldstone bosons.}
The differential geometry approach to the computation of covariant currents is specially useful when 
applied to systems with spontaneous symmetry breaking. In this case, the dynamics of NG bosons is largely determined
by the anomaly through the following prescription for the WZW effective action \cite{Zumino:1983ew}
\begin{align}
\Gamma[A,U]_{\rm WZW} =  \Gamma[\mcA]_{\rm CS} - \Gamma[\mcA^{U}]_{\rm CS},
\end{align}
where $\mcA^{U}=U^{-1}\mcA U+U^{-1}dU$ is the gauge connection transformed by $U$. The corresponding
equilibrium partition function is computed implementing the dimensional reduction to give~\cite{Manes:2018llx}
\begin{align}
W[\mcA_{0},\mbA,U]_{\rm WZW}&=i\Gamma[\mcA_{0},\mbA]_{\rm CS}-i\Gamma[\mcA_{0}^{U},\mbA^{U}]_{\rm CS}
\nonumber \\[0.2cm]
&=W[\mcA_{0},\mbA]_{\rm anom}-W[\mcA_{0},\mbA+dU U^{-1}]_{\rm anom}.
\end{align}
Here, the gauge invariant nonlocal piece of the partition function in Eq. \eqref{eq:inv+anom_decomp} cancels
between the two terms in the first line. Thus, the partition function for NG bosons can be computed
using the prescription \eqref{eq:prescription_anom}. Again, we focus our analysis to the case of the Bardeen anomaly, 
where chiral fermions couple to vector and axial-vector external gauge fields. 
If the gauge group $\mcG\times\mcG$ breaks down to its vector subroup, 
$\mcG\times\mcG\rightarrow\mcG$, the resulting WZW partition function has the structure
\begin{align}
W[\mcV_{0},\mcA_{0},\mbV,\mbA,U]_{\rm WZW}=W[\mcV_{0},\mcA_{0},\mbV,\mbA]_{\rm anom}-
W[\mcV_{0}^{U},\mcA_{0}^{U},\mbV^{U},\mbA^{U}]_{\rm anom},
\end{align}
where the transformed dimensionally-reduced gauge fields are given by
\begin{align}
\mcV_{0}^{U}&={1\over 2}(\mcV_{0}+\mcA_{0})+{1\over 2}U^{-1}(\mcV_{0}-\mcA_{0})U, \nonumber \\[0.2cm]
\mcA_{0}^{U}&={1\over 2}(\mcV_{0}+\mcA_{0})-{1\over 2}U^{-1}(\mcV_{0}-\mcA_{0})U,
\end{align}
together with
\begin{align}
\mbV^{U}&={1\over 2}(\mbV+\mbA)+{1\over 2}U^{-1}(\mbV-\mbA)U+{1\over 2}U^{-1}dU, \nonumber \\[0.2cm]
\mbA^{U}&={1\over 2}(\mbV+\mbA)-{1\over 2}U^{-1}(\mbV-\mbA)U-{1\over 2}U^{-1}dU.
\end{align}

One of the main results of Ref.~\cite{Manes:2018llx} is that the covariant current for systems with spontaneous
symmetry breaking can be computed without resorting to the WZW effective action, which usually contains 
a large number of terms. Instead, they can be obtained from the (Hodge dual) BZ current $\mcJ(\mcA)_{\rm BZ}$ 
of the unbroken theory. For the Bardeen anomaly, the vector and
axial-vector parts of the covariant currents can be obtained from the identity
\begin{align}
c_{n}\ell\widetilde{\omega}^{0}_{2n-1}(\mcV,\mcA)=-{\rm Tr\,}\Big(B_{V}\mcJ^{V}_{\rm BZ}+
B_{A}\mcJ^{A}_{\rm BZ}\Big),
\end{align}
where $\widetilde{\omega}^{0}_{2n-1}(\mcV,\mcA)$ is the Chern-Simons form preserving vector gauge transformations, 
while the operator $\ell$ is defined by its action over the gauge field and field strength forms \cite{Manes:1985df}
\begin{align}
\ell\mcV=\ell\mcA=0, \hspace*{1cm} \ell\mcF_{V}=B_{V}, \hspace*{1cm} \ell\mcF_{A}=B_{A}.
\end{align} 
In four-dimensions, the relevant Chern-Simons form $\widetilde{\omega}^{0}_{5}(\mcV,\mcA)$ is given in Eq. \eqref{eqw:omega05-tilde}. After a short calculation, 
one arrives at the following expressions for the BZ currents in terms of KK-variant vector and axial-vector gauge fields and
their fields strengths
\begin{align}
\langle\mcJ^{\mu}_{aV}\rangle_{\rm BZ}&=-{N_{c}\over 8\pi^{2}}\epsilon^{\mu\nu\alpha\beta}{\rm Tr\,}
\left[t_{a}\left(\mcV_{\nu\alpha}\mcA_{\beta}+\mcA_{\nu}\mcV_{\alpha\beta}
+{8i\over 3}\mcA_{\nu}\mcA_{\alpha}\mcA_{\beta}\right)\right], \nonumber \\[0.2cm]
\langle\mathcal{J}_{aA}^{\mu}\rangle_{\rm BZ}&=-{N_{c}\over 24\pi^{2}}\epsilon^{\mu\nu\alpha\beta}
{\rm Tr\,}\Big[t_{a}\Big(\mcA_{\nu\alpha}\mcA_{\beta}+\mcA_{\nu}\mcA_{\alpha\beta}\Big)\Big],
\label{eq:BZ_current_VA_gen}
\end{align}
where $t_{a}$ are the generators of the gauge group $\mcG$. 

Once the BZ currents are known, the right- and left-handed components of the (dual)
covariant current in the spontaneously broken symmetry phase is given by the simple relation~\cite{Manes:2018llx}
\begin{align}
\mcJ^{R}_{\mu}(\mcA_{R},\mcA_{L},U)_{\rm cov}&=\mcJ^{R}_{\mu}(\mcA_{R},\mcA_{L}^{U})_{\rm BZ}, \nonumber \\[0.2cm]
\mcJ^{L}_{\mu}(\mcA_{R},\mcA_{L},U)_{\rm cov}&=U\mcJ^{L}_{\mu}(\mcA_{R},\mcA_{L}^{U})_{\rm BZ}U^{-1},
\label{eq:JRJL_cov_SSB}
\end{align}
where the $\mcA_{R}$ and $\mcA_{L}$ gauge fields are defined in terms of $\mcV_{\mu}$ and $\mcA_{\mu}$ respectively by
\begin{align}
\mcA_{R}&=\mcV+\mcA, \nonumber \\[0.2cm]
\mcA_{L}&=\mcV-\mcA, 
\end{align}
while the corresponding components of the BZ left- and right-handed one-form  currents are 
\begin{align}
\mcJ^{R}(\mcV,\mcA)_{\rm BZ}&={1\over 2}\Big[\mcJ^{V}(\mcV,\mcA)_{\rm BZ}+\mcJ^{A}(\mcV,\mcA)_{\rm BZ}\Big],
\nonumber \\[0.2cm]
\mcJ^{L}(\mcV,\mcA)_{\rm BZ}&={1\over 2}\Big[\mcJ^{V}(\mcV,\mcA)_{\rm BZ}-\mcJ^{A}(\mcV,\mcA)_{\rm BZ}\Big]. 
\end{align}
The expressions for $\mcJ^{R}_{\rm cov}$ and $\mcJ^{L}_{\rm cov}$ have been computed
in~\cite{Manes:2018llx} using \eqref{eq:BZ_current_VA_gen}. Here we write them in flavor components 
\begin{align}
\mcJ^{R}_{a}(\mcA_R,&\mcA_L,U)_{\rm cov}=-{i\over 24\pi^{2}}{\rm Tr\,}\left\{t_{a}\left[\Big(U^{-1}\mcA_LU-\mcA_R+U_R\Big)\left(\mcF_R+\frac{1}{2}U^{-1}\mcF_LU
\right)\right.\right. \nonumber\\[0.2cm]
&+\left.\left.\left(\mcF_R+\frac{1}{2}U^{-1}\mcF_LU\right)\Big(U^{-1}\mcA_LU-\mcA_R+U_R\Big)
-\frac{1}{2}\Big(U^{-1}\mcA_LU-\mcA_R+U_R\Big)^3\right]\right\}, \nonumber \\[0.2cm]
\mcJ_{a}^{L}(\mcA_R,&\mcA_L,U)_{\rm cov}=-{i\over 24\pi^{2}}{\rm Tr\,}\left\{t_{a}\left[\Big(\mcA_L-U\mcA_R\,U^{-1}+U_L\Big)\left(\mcF_L+\frac{1}{2}U\mcF_RU^{-1}\right)\right.\right. 
\label{eq:JR_cov_SSB_explicit}\\[0.2cm] &+\left.\left.\left(\mcF_L+\frac{1}{2}U\mcF_RU^{-1}\right)\Big(\mcA_L-U\mcA_R\,U^{-1}+U_L\Big)
-\frac{1}{2}\Big(\mcA_L-U\mcA_R\,U^{-1}+U_L\Big)^3\right]\right\},
\nonumber
\end{align}
where the notation
\begin{align}
U_{R}=U^{-1}dU, \hspace*{1cm} U_{L}=dU U^{-1},
\end{align}
has been used. 
Once these left- and right-handed currents have been evaluated, the vector and axial-vector covariant dual currents are obtained by
\begin{align}
\mcJ^{V}_{a}(\mcV,\mcA,U)_{\rm cov}&=\mcJ^{R}_{a}(\mcV+\mcA,\mcV-\mcA,U)_{\rm cov}+\mcJ^{L}_{a}(\mcV+\mcA,\mcV-\mcA,U)_{\rm cov}, 
\nonumber \\[0.2cm]
\mcJ^{A}_{a}(\mcV,\mcA,U)_{\rm cov}&=\mcJ^{R}_{a}(\mcV+\mcA,\mcV-\mcA,U)_{\rm cov}-\mcJ^{L}_{a}(\mcV+\mcA,\mcV-\mcA,U)_{\rm cov}.
\label{eq:JRJL_cov_SSB_explicit}
\end{align}

\section{Two-flavor QCD fluid with unbroken chiral symmetry} 
\label{sec:2fQCD_unbroken}

We apply the formalism reviewed above first to the computation of the covariant currents and the energy-momentum tensor of 
a QCD fluid with two flavors on the static geometry~\eqref{eq:static_line_element}. 
We work in the chiral limit, where the global symmetry group is U(2)$_{L}\times$U(2)$_{R}$. 
Since the gauge group is non-Abelian, the maximal number of chemical potentials that can be consistently
introduced equals the dimension of the Cartan subalgebra \cite{Haber:1981ts,Yamada:2006rx}. 
In our case, the U(2) factors are generated by
\begin{align}
t_{0}&={1\over 2}\mathbb{1}, \hspace*{1cm}
t_{i}={1\over 2}\sigma_{i},
\label{eq:generators_U2}
\end{align}
with $\sigma_{i}$ the Pauli matrices.
Accordingly, we take background fields lying on the combination of the generators $t_0$ and $t_3$
and define the KK-invariant background
\begin{align}
V_{0}(\x)&=V_{00}(\x)t_{0}+V_{30}(\x)t_{3}, \nonumber \\[0.2cm]
V_{i}(\x)&=V_{0i}(\x)t_{0}+V_{3i}(\x)t_{3}, \nonumber \\[0.2cm]
A_{0}(\x)&= A_{00}(\x)t_{0},  
\label{eq:2fQCD_background_t0t3}\\[0.2cm]
 A_{i}(\x)&=0. \nonumber
\end{align} 
Furthermore, we take a nonzero {\em constant} value for $ A_{00}$, which is related to the axial
chemical potential $\mu_{5}$ controlling chiral imbalance by\footnote{This definition of the axial chemical potential differ by a factor of $2$ from
the one used in \cite{Jensen:2013kka,Manes:2018llx}, where this quantity was introduced through the identity 
$\mcA_{\mu}= A_{i}\delta^{i}_{\mu}-\mu_{5}u_{\mu}\mathbb{1}$. To compare with the results of these references, one should replace
$\mu_{5}\rightarrow 2\mu_{5}$ in our expressions.}
\begin{align}
\mu_{5}=e^{-\sigma} A_{00}.
\label{eq:A00mu5}
\end{align}
The other two chemical potentials $\mu_{0}$ and $\mu_{3}$ are associated with the time-components of the vector gauge fields according to
\begin{align}
\mu_{0}&=e^{-\sigma} V_{00}, \nonumber \\[0.2cm]
\mu_{3}&=e^{-\sigma} V_{30},
\label{eq:mu0+mu3_defs}
\end{align}
and respectively control baryon number and isospin imbalance. 
By choosing $ A_{00}$ to be constant, we are focusing on corrections to the 
covariant currents and energy-momentum tensor that contain a  
single derivative of the vector background fields. In addition, as in~\cite{Gatto:2011wc,Andrianov:2013qta},  
we have not included terms in $ A_{0}$ proportional to $t_{3}$, which would contribute to the axial-vector current. 

Using Eq. \eqref{eq:general_decomp_Afield}, the gauge fields can be written in a formally Lorentz covariant form in terms
of the chemical potentials and the four-velocity. For
$\mcA_{0\mu}$, since the spatial components are zero, the gauge field itself is proportional to the four-velocity vector
\begin{align}
\mcA_{0\mu}=-e^{-\sigma} A_{00}u_{\mu}=-\mu_{5}u_{\mu}.
\label{eq:cov_form_mbhA_mu}
\end{align}
The associated field strength can then be written as
\begin{align}
\mcA_{0\mu\nu}=-\mu_{5}\big(\partial_{\mu}u_{\nu}-\partial_{\nu}u_{\mu}\big)
+\big(u_{\mu}\partial_{\nu}\mu_{5}-u_{\nu}\partial_{\mu}\mu_{5}\big),
\end{align}
where we identify two terms respectively
related to the vorticity tensor and the spacetime
variations of the chemical potential $\mu_{5}$ governing chiral imbalance. 
For the vector gauge fields $\mcV_{a\mu}$, we find ($a=0,3$)
\begin{align}
\mcV_{a\mu}=-e^{-\sigma} V_{a0}u_{\mu}+\delta^{i}_{\mu} V_{ai}
=-\mu_{a}u_{\mu}+\delta^{i}_{\mu} V_{ai},
\end{align}
whose field strength tensor, besides the vorticity and chemical potential gradient contributions pointed out in the axial-vector field,
has an extra piece  
\begin{align}
\mcV_{a\mu\nu}&=-\mu_{a}\big(\partial_{\mu}u_{\nu}-\partial_{\nu}u_{\mu}\big)
+\big(u_{\mu}\partial_{\nu}\mu_{a}-u_{\nu}\partial_{\mu}\mu_{a}\big) 
+\delta^{i}_{\mu}\delta^{j}_{\nu} V_{aij},
\end{align}
with $ V_{aij}=\partial_{i} V_{aj}-\partial_{j} V_{ai}$ the field strength of the KK-invariant components of the
Abelian background vector gauge field.

\subsection{Partition function and gauge currents}

At leading (zeroth) order in the derivative expansion, the partition function can be written in terms of the pressure $P_{0}$, which depends on the
temperature and the chemical potentials $\mu_{0}$ and $\mu_{3}$ defined in Eq.~\eqref{eq:mu0+mu3_defs}, namely
\begin{align}
W={1\over T_{0}}\int d^{3}x\,\sqrt{g}e^{\sigma}P_{0}(\mu_{0},\mu_{3},T).
\label{eq:leading_order_noGB}
\end{align}
The first order correction to this result is given by 
the anomalous equilibrium partition function computed
in \cite{Manes:2018llx} in the language of differential forms. For a general background, the same expression written 
in components gives 
\begin{align}
W[ V_0, A_0, V_{i}, A_{i}&]_{\rm anom}^{{\rm U}(2)\times{\rm U}(2)}
={N_{c}\over 8\pi^{2} T_0}\int\limits_{S^{3}}d^{3}x\sqrt{g}\epsilon^{ijk} \left\{   
\big({\rm Tr\,} V_{0}\big){\rm Tr\,}\left( V_{ij} A_{k}+{4i\over 3} A_{i} A_{j} A_{k}\right)
\right.\nonumber \\[0.2cm]
&+\big({\rm Tr\,} V_{ij}\big){\rm Tr\,}\big( V_{0} A_{k}\big)
+\Big[{\rm Tr\,}\big( V_{0} V_{ij}\big) 
-({\rm Tr\,} V_{0})({\rm Tr\,} V_{ij})\Big]({\rm Tr\,} A_{k})
\nonumber \\[0.2cm]
&+{1\over 3}\big({\rm Tr\,} A_{ij}\big){\rm Tr\,}\big( A_{0} A_{k}\big)
+{1\over 3}\Big[{\rm Tr\,}\big( A_{0} A_{ij}\big)
-\big({\rm Tr\,} A_{0}\big)\big({\rm Tr\,} V_{ij}\big)\Big]\big({\rm Tr\,} A_{k}\big) \nonumber \\[0.2cm]
&+{1\over 3}\big({\rm Tr\,} A_{0}\big){\rm Tr\,}\big( A_{ij} A_{k}\big) -{4\over 3}\big({\rm Tr\,} A_{i}\big){\rm Tr\,}\big( V_{0} A_{j} A_{k}\big)
\label{eq:W_anom_VA_U2xU2} \\[0.2cm]
&-{1\over 2}f_{ij}\Big[({\rm Tr\,} V_{0})^{2}-{\rm Tr\,} V_{0}^{2}\Big]({\rm Tr\,} A_{k})
-{1\over 6}f_{ij}\Big[\big({\rm Tr\,} A_{0}\big)^{2}-{\rm Tr\,} A_{0}^{2}\Big]\big({\rm Tr\,} A_{k}\big)
\nonumber \\[0.2cm]
&+f_{ij}\big({\rm Tr\,} V_{0}\big)
{\rm Tr\,}\big( V_{0} A_{k}\big)+{1\over 3}f_{ij}\big({\rm Tr\,} A_{0}\big){\rm Tr\,}\big( A_{0} A_{k}\big) \bigg\}
\nonumber \\[0.2cm]
&+{1\over 2}C_{2}T_{0}\int\limits_{S^{3}} d^{3}\x \sqrt{g} \, \epsilon^{ijk}f_{ij}\big({\rm Tr\,} A_{k}\big).
\nonumber
\end{align} 
Here $\epsilon^{ijk}$ represents the Levi-Civita pseudotensor, normalized according to $\epsilon^{123}=1/\sqrt{g}$. 
The last term in the anomalous effective action 
mixes the axial-vector gauge field with the KK vector field and is responsible for the corrections to the chiral vortical effect 
quadratic in the temperature. Various analyses show \cite{Landsteiner:2011cp,Golkar:2012kb,Jensen:2012kj} that 
the coefficient $C_{2}$ is related to the value 
of the mixed gauge-gravitational anomaly and is not computable using the differential geometry methods used here. 
We stress that the functional given in Eq. \eqref{eq:W_anom_VA_U2xU2} only contains KK-invariant gauge fields, while its explicit 
dependence on the metric function $a_{i}$ comes through the Abelian field strength 
\begin{align}
f_{ij}\equiv \partial_{i}a_{j}-\partial_{j}a_{i}.
\end{align}
Thus, the equilibrium partition function is explicitly invariant under KK gauge transformations.

The covariant vector and axial-vector currents can be computed using Eqs. \eqref{eq:currentsJ_vJ_a}
and \eqref{eq:currentsJ_vJ_a_0comp},
particularized to the background \eqref{eq:2fQCD_background_t0t3}.
Expressing them in terms of the four-velocity $u^{\mu}$, 
the chemical potentials \eqref{eq:A00mu5} and~\eqref{eq:mu0+mu3_defs}, and the field strength 
components $\mcV_{\mu\nu}=\mcV_{a\mu\nu}t_{a}$ of the KK-variant fields [see~\eqref{eq:FV_FA_nonKK_defs}], we arrive at
\begin{align}
\langle \mathcal{J}_{0V}^{\mu}\rangle_{\rm cov}&={N_{c}\over 16\pi^{2}}\mu_{5}\epsilon^{\mu\nu\alpha\beta}u_{\nu}\mcV_{0\alpha\beta}, \nonumber \\[0.2cm]
\langle \mathcal{J}_{3V}^{\mu}\rangle_{\rm cov}&={N_{c}\over 16\pi^{2}}\mu_{5}\epsilon^{\mu\nu\alpha\beta}u_{\nu}\mcV_{3\alpha\beta}, \nonumber \\[0.2cm]
\langle \mathcal{J}_{0A}^{\mu}\rangle_{\rm cov}&={N_{c}\over 16\pi^{2}}\epsilon^{\mu\nu\alpha\beta}u_{\nu}\Big[\mu_{0}\mcV_{0\alpha\beta}
+\mu_{3}\mcV_{3\alpha\beta}+\big(\mu_{0}^{2}+\mu_{3}^{2}-\mu_{5}^{2}\big)\partial_{\alpha}u_{\beta}\Big], 
\label{eq:JV_JA_2f}\\[0.2cm]
\langle \mathcal{J}_{3A}^{\mu}\rangle_{\rm cov}&={N_{c}\over 16\pi^{2}}\epsilon^{\mu\nu\alpha\beta}u_{\nu}\Big(\mu_{3}\mcV_{0\alpha\beta}+\mu_{0}\mcV_{3\alpha\beta}
+2\mu_{0}\mu_{3}\partial_{\alpha}u_{\beta}\Big),
\nonumber
\end{align}
where we have followed the notation introduced in Eq. \eqref{eq:current_ifactor}.
In these expressions and in the following, the four-dimensional Levi-Civita pseudotensor is normalized such 
that\footnote{In this work, we depart from the conventions of
Ref. \cite{Nakahara:2003nw}, where the Levi-Civita tensor is defined with an additional power of $\sqrt{-G}$.}
\begin{align}
\epsilon^{0123}={1\over \sqrt{-G}}={e^{-\sigma}\over \sqrt{g}} \hspace*{1cm} \Longrightarrow \hspace*{1cm}
\epsilon^{0ijk}=e^{-\sigma}\epsilon^{ijk},
\label{eq:normalization_epsilon}
\end{align} 
with $G$ the determinant of the static metric \eqref{eq:static_line_element} and $\epsilon^{ijk}$ the three-dimensional
antisymmetric pseudotensor normalized as $\epsilon^{123}=1/\sqrt{g}$ [cf. Eq. \eqref{eq:W_anom_VA_U2xU2}].  

The second quantity to be conside\ is 
the anomalous contribution to the energy-momentum tensor, codified in the vector $q^{\mu}$ defined by
\begin{align}
T^{\mu\nu}=u^{\mu}q^{\nu}+u^{\nu}q^{\mu}.
\end{align}
This can be evaluated from Eq. \eqref{eq:EM_tensor_Bardeen} and takes the form
\begin{align}
q^{\mu}&={N_{c}\over 16\pi^{2}}\epsilon^{\mu\nu\alpha\beta}u_{\nu}\mu_{5}\left[\mu_{0}\mcV_{0\alpha\beta}+\mu_{3}\mcV_{3\alpha\beta}
+\left(\mu_{0}^{2}+\mu_{3}^{2}-{1\over 3}\mu_{5}^{2}\right)\partial_{\alpha}u_{\beta}\right].
\label{eq:qmu_2f}
\end{align}
These Abelianized expressions match the general structure for the anomalous currents 
derived in Ref.~\cite{Neiman:2010zi}. 
It should be pointed out 
that the vectors in Eqs. \eqref{eq:JV_JA_2f} and \eqref{eq:qmu_2f} are transverse to $u^{\mu}$, as it follows from the general structure
of the covariant currents found in Eq. \eqref{eq:currentsJ_vJ_a_0comp}. As we will see 
in Sec. \ref{sec:corrections_LO}, this property is broken in the presence of NG bosons, where the covariant currents develop a longitudinal 
component.

Inspecting the different terms in the currents and energy-momentum tensor given in Eqs. \eqref{eq:JV_JA_2f} and
\eqref{eq:qmu_2f}, we find that the coupling to the external gauge fields comes through the magnetic components of the 
non-Abelian vector field strength ($a=0,3$)
\begin{align}
\mcB^{\mu}_{a}\equiv {1\over 2}\epsilon^{\mu\nu\alpha\beta}u_{\nu}\mcV_{a\alpha\beta}
&=\bigg(-a_{j}\epsilon^{jk\ell}\partial_{k}\mcV_{a\ell},\epsilon^{ijk}\Big(\partial_{j}\mcV_{ak}+a_{j}\partial_{k}\mcV_{a0}\Big)
\bigg) \nonumber \\[0.2cm]
&=\left(-a_{j}\epsilon^{jk\ell}\left(\partial_{k} V_{a\ell}
+{1\over 2} V_{a0}f_{k\ell}\right),\epsilon^{ijk}\left(\partial_{j} V_{ak}+{1\over 2} V_{a0}f_{jk}\right)
\right),
\label{eq:magnetic_field}
\end{align}
where the second line is obtained by applying the decomposition \eqref{eq:general_decomp_Afield} to the definition of the magnetic
field.
The explicit dependence on the four-velocity $u_{\mu}$, on the other hand,
is codified in terms of the vorticity vector
\begin{align}
\omega^{\mu}\equiv {1\over 2}\epsilon^{\mu\nu\alpha\beta}u_{\nu}\partial_{\alpha}u_{\beta}
={1\over 4}e^{\sigma}\Big(\epsilon^{jk\ell}a_{j}f_{k\ell},-\epsilon^{ijk}f_{jk}\Big).
\label{eq:vorticity_field}
\end{align}
Having identified these transverse structures, we write the constitutive relations in covariant form
\begin{align}
\langle \mathcal{J}^{\mu}_{aV}\rangle_{\rm cov} &=\xi_{V,B}^{ab}\mcB^{\mu}_{b}+\xi_{V}^{a}\omega^{\mu}, 
\nonumber \\[0.2cm]
\langle \mathcal{J}^{\mu}_{aA}\rangle_{\rm cov} &=\xi_{A,B}^{ab}\mcB^{\mu}_{b}+\xi_{A}^{a}\omega^{\mu}, 
\label{eq:constitutive_unbroken}\\[0.2cm]
q^{\mu}&=\xi^{a}_{\epsilon,B}\mcB^{\mu}_{a}+\xi_{\epsilon,V}\omega^{\mu},
\nonumber
\end{align}
so the corresponding out-of-equilibrium anomalous transport coefficients
associated with the chiral magnetic, the chiral vortical, and the chiral separation effects
can be read from Eqs. \eqref{eq:JV_JA_2f} and~\eqref{eq:qmu_2f}, to give
\begin{align}
\xi_{V,B}^{ab}&={N_{c}\over 8\pi^{2}}\mu_{5}\delta^{ab}, \hspace*{3.4cm}
\xi_{V}^{a}=0, \nonumber \\[0.2cm]
\xi_{A,B}^{ab}&={N_{c}\over 8\pi^{2}}\Big[(\mu_{0}-\mu_{3})\delta^{ab}+\mu_{3}\Big]
, \hspace*{1cm} \xi_{A}^{a}={N_{c}\over 8\pi^{2}}\Big[
\Big(\mu_{0}^{2}+\mu_{3}^{2}-\mu_{5}^{2}\Big)\delta^{a0}+2\mu_{0}\mu_{3}\delta^{a3}\Big]
,
\label{eq:transport_coeff_unbroken}\\[0.2cm]
\xi^{a}_{\epsilon,B}&={N_{c}\over 8\pi^{2}}\mu_{5}\Big(\mu_{0}\delta^{a0}+\mu_{3}\delta^{a3}\Big)
, \hspace*{0.9cm} \xi_{\epsilon,V}={N_{c}\over 8\pi^{2}}\mu_{5}\left(\mu_{0}^{2}+\mu_{3}^{2}-{1\over 3}\mu_{5}^{2}\right).
\nonumber
\end{align}

\subsection{Baryon, isospin, and electromagnetic currents}

We define now the vector baryonic and isospin currents as those associated with the generators $t_{0}$ and $t_{3}$, which
properly normalized read
\begin{align}
\mathcal{J}_{\rm bar}^{\mu}&={2\over 3}\overline{\Psi}\gamma^{\mu}t_{0}\Psi, 
\label{eq:bar_iso_currents}\\[0.2cm]
\mathcal{J}_{\rm iso}^{\mu}&=\overline{\Psi}\gamma^{\mu}t_{3}\Psi, \nonumber
\end{align}
where $\Psi$ is a flavor doublet of Dirac spinors made out of the two light quark flavors
\begin{align}
\Psi=\left(
\begin{array}{c}
u \\
d
\end{array}
\right),
\end{align}
with each spinor being decomposed into its two chiralities according to
\begin{align} 
u=\left(
\begin{array}{c}
u_{L} \\
u_{R}
\end{array}
\right), \hspace*{0.5cm} \hspace*{0.2cm}
d=\left(
\begin{array}{c}
d_{L} \\
d_{R}
\end{array}
\right).
\end{align}

Electromagnetism, on the other hand, is identified with the U(1)$_{V}$ subgroup generated by the charge matrix $Q$ 
defined by the Gell-Mann--Nishijima formula
\begin{align}
Q\equiv {1\over 3}t_{0}+t_{3}=\left( 
\begin{array}{cc}
{2\over 3} &  0 \\
0 &  -{1\over 3} 
\end{array} 
\right),
\label{eq:charge_matrix}
\end{align}
where charges are expressed in units of the elementary charge $e$. In terms of $Q$, 
the electromagnetic current is now given by
\begin{align}
\mathcal{J}_{\rm em}^{\mu}&\equiv e\overline{\Psi}\gamma^{\mu}Q\Psi={e\over 3}\overline{\Psi}\gamma^{\mu}t_{0}\Psi+e\overline{\Psi}\gamma^{\mu} 
t_{3}\Psi,
\label{eq:em_currents}
\end{align}
where we have introduced the explicit dependence on $e$.

Since the electromagnetic field is the only propagating gauge field in the theory, it is
convenient to take $Q$ as one of the elements of the basis of the Cartan subalgebra of the vector factors $\mbox{SU(2)$_{V}\times$U(1)$_{V}$}$,
instead of the U(1)$_{B}$ generator $t_{0}$. Then, the vector background field $ V_{\mu}$ can be written as
\begin{align}
 V_{\mu}= V_{0\mu}t_{0}+ V_{3\mu}t_{3}= 3 V_{0\mu}Q+\Big( V_{3\mu}-3 V_{0\mu}\Big)t_{3}.
\end{align}
We thus identify the electromagnetic potential as the gauge field coupling to 
$Q$, while the component along the generator of the Cartan subalgebra of SU(2)$_{V}$ is set to zero, since it is just an auxiliary field
which does not contain
propagating degrees of freedom at low energies
\begin{align}
\left.
\begin{array}{rl}
3 V_{0\mu}& \equiv e\mathbb{V}_{\mu}  \\[0.2cm]
 V_{3\mu}-3 V_{0\mu}& =0 
\end{array}
\right\} \hspace*{1cm}
\Longrightarrow \hspace*{1cm}  V_{3\mu}=3 V_{0\mu}=e\mathbb{V}_{\mu}.
\label{eq:EM_field_def}
\end{align}
This in turn implies the following relation between the isospin and baryonic chemical potentials
\begin{align}
\mu_{0}={1\over 3}\mu_{3}={e\over 3}e^{-\sigma}\mathbb{V}_{0}.
\label{eq:relationmu0mu3V0}
\end{align}

At the quantum level, the expectation values of the consistent electromagnetic, baryonic, and isospin currents can be written in terms of the values
of the consistent vector currents $\langle\mathcal{J}^{\mu}_{aV}\rangle_{\rm cons}$ (with $a=0,3$), obtained by varying the quantum effective action 
\begin{align}
\langle \mathcal{J}^{\mu}_{\rm em}\rangle_{\rm cons}&={e\over 3}\langle \mathcal{J}^{\mu}_{0V}\rangle_{\rm cons}
+e\langle \mathcal{J}^{\mu}_{3V}\rangle_{\rm cons}, \nonumber \\[0.2cm]
\langle \mathcal{J}^{\mu}_{\rm bar}\rangle_{\rm cons} &= {2\over 3}\langle \mathcal{J}_{0V}^{\mu}\rangle_{\rm cons},
\label{eq:J_em,bar,iso_def} \\[0.2cm]
\langle \mathcal{J}^{\mu}_{\rm iso}\rangle_{\rm cons} &= \langle \mathcal{J}_{3V}^{\mu}\rangle_{\rm cons}.
\nonumber
\end{align}
The same pattern is followed by the BZ terms of each current and, as a consequence, the same relations are satisfied by the corresponding
covariant currents. Given these expressions, it is clear that the currents defined in Eqs. \eqref{eq:bar_iso_currents} 
and \eqref{eq:em_currents} satisfy 
the Gell-Mann--Nishijima relation
also at the quantum level
\begin{align}
\langle \mathcal{J}^{\mu}_{\rm em}\rangle={e\over 2}\langle \mathcal{J}^{\mu}_{\rm bar}\rangle+e\langle \mathcal{J}^{\mu}_{\rm iso}\rangle.
\end{align}

We can use now the constitutive relations \eqref{eq:constitutive_unbroken} and transport coefficients \eqref{eq:transport_coeff_unbroken} 
to find the constitutive relations for the covariant electromagnetic current
\begin{align}
\langle \mathcal{J}^{\mu}_{\rm em}\rangle_{\rm cov}&={e\over 3}\langle\mathcal{J}^{\mu}_{0V}\rangle_{\rm cov}+e\langle\mathcal{J}_{3V}^{\mu}\rangle_{\rm cov} \nonumber \\[0.2cm]
&={eN_{c}\over 24\pi^{2}}\mu_{5}\Big(\mcB^{\mu}_{0}+3\mcB^{\mu}_{3}\Big).
\label{eq:constitutive_unbroken1}
\end{align}
To recast the term inside the bracket in terms of the physical electromagnetic field, we recall Eq. \eqref{eq:EM_field_def}, from where the following 
expression of the KK-variant components of the electromagnetic potential $\mcV_{\mu}$ follows
\begin{align}
(\mcV_{0},\mcV_{i})\equiv (\mathbb{V}_{0},\mathbb{V}_{i}+\mathbb{V}_{0}a_{i})
\hspace*{1cm} \Longrightarrow \hspace*{1cm} 
\left\{
\begin{array}{rl}
\mcE_{3}^{\mu}=3\mcE_{0}^{\mu}&=e\mcE^{\mu} \\[0.2cm]
\mcB_{3}^{\mu}=3\mcB_{0}^{\mu}&=e\mcB^{\mu}
\end{array}
\right.
\label{eq:cov_electric_field_vs_E3E0}
\end{align}
where the KK-variant electric and magnetic fields are defined in the usual way from the corresponding field strength
\begin{align}
\mcE_{\mu}&\equiv \mcV_{\mu\nu}u^{\nu}, \nonumber \\[0.2cm]
\mcB^{\mu}&\equiv {1\over 2}\epsilon^{\mu\nu\alpha\beta}u_{\nu}\mcV_{\alpha\beta}.
\label{eq:magneticfield_KK-variant_def}
\end{align}
With this, the constitutive equation for the electromagnetic current in Eq. \eqref{eq:constitutive_unbroken1} reads
\begin{align}
\langle \mathcal{J}^{\mu}_{\rm em}\rangle_{\rm cov}&={5e^{2}N_{c}\over 36\pi^{2}}\mu_{5}\mcB^{\mu},
\label{eq:final_constitutive_unbroken}
\end{align}
which also gives the transport coefficient associated with the chiral magnetic effect. On the other hand, the vanishing of the 
coefficient $\xi_{V}^{a}$ implies the absence of a chiral vortical effect in the
vector currents of the unbroken theory.

Alternatively, we can express the current in terms of the KK-invariant magnetic field defined from the electromagnetic potential 
$\mathbb{V}_{\mu}$
\begin{align}
\mathbb{B}^{\mu}\equiv\mathcal{B}^{\mu}+{6\over e}\mu_{0}\omega^{\mu}
=\Big(-a_{j}\epsilon^{jk\ell}\partial_{k}\mathbb{V}_{\ell},\epsilon^{ijk}\partial_{j}\mathbb{V}_{k}\Big),
\label{eq:mag_fieldbb_def}
\end{align}
to read
\begin{align}
\langle \mathcal{J}^{\mu}_{\rm em}\rangle_{\rm cov}&={5e^{2}N_{c}\over 36\pi^{2}}\mu_{5}\mathbb{B}^{\mu}-{5eN_{c}\over 6\pi^{2}}
\mu_{0}\mu_{5}\omega^{\mu}.
\label{eq:final_constitutive_unbroken_physmag}
\end{align}
Thus, once expressed in terms of the physical fields, a contribution to the chiral vortical conductivity emerges, together with the one
associated with the chiral magnetic effect. This is similar to the results  
found in~\cite{Landsteiner:2012kd} for an Abelian U(1)$_{V}\times$U(1)$_{A}$ theory.

The cancellation leading to our result $\xi_{V}^{a}=0$ in \eqref{eq:transport_coeff_unbroken} is a direct consequence of having 
considered the flavor group U(2)$\times$U(2) and, to our knowledge, has not been noticed before in the literature. In a sense, this means
that when expressed in terms of the KK-invariant fields the chiral vortical effect in Eq. \eqref{eq:final_constitutive_unbroken_physmag}
is geometrically determined, since its value is encoded in the relation between KK-variant and KK-invariant gauge fields 
components~\eqref{eq:def_ttgaugefield_gen}. This state of affairs contrasts with the U(1)$\times$U(1) case, 
studied in \cite{Landsteiner:2012kd,Jensen:2013vta}, where such cancellation does not take place. 

\section{Pion partition function of two-flavor QCD in the chiral limit}
\label{sec:pion_part_funct}

Let us allow now for the possibility of spontaneous chiral symmetry breaking in the theory studied in the previous section.
The first thing to be taken into account is that 
the axial symmetry U(1)$_{A}$ is already broken by nonperturbative effects and therefore not realized as an invariance of the
theory. As a consequence, we focus on the spontaneous breaking of the surviving subgroups SU(2)$_{L}\times$SU(2)$_{R}\times$U(1)$_{V}$ 
down to its vector-like factors SU(2)$_{V}\times$U(1)$_{V}$. In the process, three NG bosons (the three pions) 
appear, corresponding to
the broken generators of SU(2)$_{A}$. No NG boson is associated with the U(1)$_{A}$ factor since, as mentioned,
this symmetry is absent from the start. 

These NG bosons are codified in the 
matrix $U$, which can be parametrized
using the Pauli matrices $\sigma_{a}$ as
\begin{align}
U=\exp\left(i\sum_{a=1}^{3}\zeta_{a}\sigma_{a}\right)
\hspace*{0.5cm} \mbox{with} \hspace*{0.5cm}
\sum_{a=1}^{3} \zeta_{a} \sigma_{a}={\sqrt{2}\over f_\pi} 
\left(
\begin{array}{cc}
\tfrac{1}{\sqrt{2}} \pi^0 & \pi^+ \\
\pi^- & -\tfrac{1}{\sqrt{2}} \pi^0
\end{array} \right),  
\end{align}
with $\pi^{0}, \pi^{\pm}$ the three pion fields and $f_{\pi}\approx 92\mbox{ MeV}$ the pion decay constant. 
In order to secure the invariance of the effective action under the gauge transformations of the electromagnetic potential introduced
in Eq. \eqref{eq:EM_field_def}, 
we need the NG field to transform 
according to
\begin{align}
\delta_{\chi} \mathbb{V}_{i}(\x)=\partial_{i}\chi(\x) \hspace*{1cm} \Longrightarrow \hspace*{1cm}
\delta_{\chi} U(\x)=-\chi(\x)[Q,U(\x)],
\end{align}
which gives the correct assignment of charges to the three pions. 

Having zero baryon number, pions do not couple to the gauge field associated with the baryon number factor U(1)$_{B}$ generated by $t_{0}$,
although they do couple to the gauge field along the isospin generator $t_{3}$. The
covariant derivative acting on the NG boson field reads
\begin{align}
\xi_{\mu}\equiv D_{\mu}U=\partial_{\mu}U-i\mcV_{3\mu}[t_{3},U],
\label{eq:cov_derivative_xi}
\end{align}
where we see how $\mcV_{0\mu}t_{0}$ drops out of the
commutator. 
With this in mind, 
the action at lowest order in derivatives can be written as
\begin{align}
W&={1\over T_{0}}\int d^{3}x\,\sqrt{g}e^{\sigma}P(\mu_{0},\mu_{3},T,\mcV_{3\mu},U) \nonumber \\[0.2cm]
&={1\over T_{0}}\int d^{3}x\, \sqrt{g}e^{\sigma}\Big[P_{0}(\mu_{0}, \mu_{3},T)+\mathscr{L}\Big],
\label{eq:W_leading_orderSSB}
\end{align}
where $P_0$ is the pressure in the absence of NG bosons, already introduced in Eq.~\eqref{eq:leading_order_noGB}.
The second term  $\mathscr{L}$ contains all dependence on the pions and
has the following expression in terms of KK-invariant fields 
\begin{align}
\mathscr{L}&={f_{\pi}^{2}\over 4}G^{\mu\nu}{\rm Tr\,}\Big[D_{\mu}U(D_{\nu}U)^{\dagger}\Big]
\nonumber \\[0.2cm]
&={f_{\pi}^{2} \over 4}\left[e^{-2\sigma}{\rm Tr\,}\big(\xi_{0}\xi_{0}^{\dagger}\big)
-g^{ij}{\rm Tr\,}\big(\Phi_{i}\Phi_{j}^{\dagger}\big)\right], 
\label{eq:general_lagrangian_pions_firstorder}
\end{align}
where $\Phi_{i}$ are defined by
\begin{align}
\Phi_{i}=\partial_{i}U-i V_{3i}[t_{3}, U].
\end{align}
Notice that in this expression $Q$ can replace $t_{3}$ inside the commutator, whereas from Eq. \eqref{eq:EM_field_def} we know that
$ V_{3i}=e\mathbb{V}_{i}$. Similarly, since $ V_{30}=e\mathbb{V}_{0}$, we find that $\xi_{0}=-ie\mathbb{V}_{0}[Q,U]$. 
As a consequence, the leading order Lagrangian \eqref{eq:general_lagrangian_pions_firstorder}
governs the electromagnetic coupling of the two charged pions $\pi^{\pm}$. In the case of the neutral pion, 
the analogous coupling is mediated
by the anomaly and only appears at the next order in the derivative expansion, as we will see below.

At first order in derivatives, the correction to the partition function is given by the corresponding WZW action in the background \eqref{eq:2fQCD_background_t0t3}. This quantity 
was computed in~\cite{Manes:2018llx} and in components reads
\begin{align}
W[ V_{0\mu},& V_{3\mu}, A_{00},U]_{\rm WZW}^{{\rm U}(2)\times{\rm U}(2)}
={1\over 8\pi^{2} T_0}\int\limits_{S^{3}}d^{3}x\sqrt{g}\,\epsilon^{ijk} \left\{ 
-{1\over 2} V_{00} V_{3i}\partial_{j}{\rm Tr\,}\Big[(R_{k}+L_{k})Q\Big]  
\right. \nonumber \\[0.2cm]
&+{i\over 6} V_{00}{\rm Tr\,}\big(L_{i}L_{j}L_{k}\big)
+{1\over 2}\left( V_{00}\partial_{i} V_{3j}+ V_{30}\partial_{i} V_{0j}+{1\over 2}f_{ij} V_{00} V_{30}\right){\rm Tr\,}\Big[\big(R_{k}+L_{k}\big)Q\Big]
\nonumber \\[0.2cm]
&+{1\over 6} A_{00}\left(\partial_{i} V_{3j}+{1\over 2}f_{ij} V_{30}\right){\rm Tr\,}\Big[(R_{k}-L_{k})Q\Big] 
\label{eq:general_SSB_U2xU2} \\[0.2cm]
&\left.-{1\over 3} A_{00}\left(\partial_{i} V_{3j}+{1\over 2}f_{ij}  V_{30} \right) V_{3k}{\rm Tr\,}\Big[Q\Big(Q-U^{-1}QU\Big)\Big]\right\}.
\nonumber
\end{align}
Here we have introduced the notation
\begin{align}
R_{i}&\equiv iU^{-1}\partial_{i}U, \nonumber \\[0.2cm]
L_{i}&\equiv i\partial_{i}U U^{-1}.
\end{align}
Notice that since $U$ takes values on SU(2), the generator $t_{3}$ can be interchanged with the charge matrix $Q$  
inside the traces in Eq. \eqref{eq:general_SSB_U2xU2}.

We can now expand the equilibrium partition function \eqref{eq:general_SSB_U2xU2} in powers of the pion fields.
Keeping terms with just one derivative and up to two pion fields, the basics identities to be used are
\begin{align}
{\rm Tr\,}\Big[(R_{i}+L_{i})Q\Big]&=-{2\over f_{\pi}}\partial_{i}\pi^{0}+\mathcal{O}(\pi^{3}), \nonumber \\[0.2cm]
{\rm Tr\,}\Big[(R_{i}-L_{i})Q\Big]+2\mathbb{V}_{i}{\rm Tr\,}\Big[\Big(U^{-1}QU-Q\Big)Q\Big]&=  
{2i\over f_{\pi}^{2}}\Big(\partial_{i}\pi^{-}\pi^{+}-\pi^{-}\partial_{i}\pi^{+}\Big) 
\label{eq:GB_expansion_pions}\\[0.2cm]   
&-{4\over f_{\pi}^{2}}\pi^{-}\pi^{+}\mathbb{V}_{i}+\mathcal{O}(\pi^{4}).
\nonumber
\end{align}
Plugging these expansions into Eq. \eqref{eq:general_SSB_U2xU2}, we find 
the following partition function in
terms of the pion and electromagnetic field
\begin{align}
W[\mathbb{V}_{\mu},\pi^{0},\pi^{\pm}]_{\rm WZW}
&=-\int\limits_{S^{3}}d^{3}x\sqrt{g}\, \left[{e^{2}N_{c}\over 12\pi^{2} f_{\pi}T_{0}}
\mathbb{V}_{0}\partial_{k}\pi^{0}\left(\mathbb{B}^{k}+{1\over 4}\epsilon^{ijk}f_{ij}\mathbb{V}_{0}\right)
\right.
\label{eq:pion_eff_action_V0}\\[0.2cm]
&+\left.{i\mu_{5}eN_{c}\over 24\pi^{2}f_{\pi}^{2}T}\Big(
\pi^{-}\partial_{k}\pi^{+}-\pi^{+}\partial_{k}\pi^{-}-2ie\pi^{+}\pi^{-}\mathbb{V}_{k}\Big)
\left(\mathbb{B}^{k}+{1\over 2}\epsilon^{ijk}f_{ij}\mathbb{V}_{0}\right)\right],
\nonumber
\end{align}
where the magnetic field is defined as usual by
$\mathbb{B}^{i}=\epsilon^{ijk}\partial_{j}\mathbb{V}_{k}$ [cf. Eq. \eqref{eq:mag_fieldbb_def}].
In addition, we have used Eq. \eqref{eq:A00mu5} to identify the chemical potential governing chiral imbalance, as well as the 
local temperature $T(\x)=T_{0}e^{-\sigma(\x)}$. 
While the first term gives the anomaly-mediated coupling of the neutral pion to the electromagnetic field, corrected by the 
effect of the curved background, the second one agrees with the form of the parity-odd couplings 
obtained from the WZW effective action in systems with chiral imbalance, as shown in~\cite{Witten:1983tw,Andrianov:2017ely}.

To compare with existing results in the literature, it is convenient to go back to \eqref{eq:general_SSB_U2xU2} and recast it in terms of the 
electromagnetic potential $\mathbb{V}_{\mu}$ as
\begin{align}
W[\mathbb{V}_{\mu},&U]_{\rm WZW}^{{\rm U}(2)\times{\rm U}(2)}
=\int\limits_{S^{3}}d^{3}x\sqrt{g}\,\epsilon^{ijk} \bigg( g_{1}(\nu)
\partial_{i}\mathbb{V}_{j}{\rm Tr\,}\Big[Q\big(R_{k}+L_{k}\big)\Big]
\label{eq:WZW_2f_eff_EMfield} \\[0.2cm]
&+g_{2}(\nu)\left\{
-{1\over 2}\mathbb{V}_{i}{\rm Tr\,}\Big[Q\partial_{j}(R_{k}+L_{k})\Big]
+{i\over 6e}{\rm Tr\,}\big(L_{i}L_{j}L_{k}\big)\right\} 
+{T_{0}\over 2}g_{3}(\nu)f_{ij}{\rm Tr\,}\Big[Q\big(R_{k}+L_{k}\big)\Big]\nonumber \\[0.2cm]
&\left.+g_{4}(\nu_{5})\left(\partial_{i}\mathbb{V}_{j}+{1\over 2}f_{ij}\mathbb{V}_{0}\right)\left\{ 
2\mathbb{V}_{k}{\rm Tr\,}\Big[Q\Big(Q-U^{-1}QU\Big)\Big]-{1\over e}{\rm Tr\,}\Big[Q(R_{k}-L_{k})\Big]\right\}\right),
\nonumber
\end{align}
where we have written
\begin{align}
\nu(\x)&\equiv {\mu(\x)\over T(\x)}={\mathbb{V}_{0}(\x)\over T_{0}}, \nonumber \\[0.2cm]
\nu_{5}(\x)&\equiv {\mu_{5}(\x)\over T(\x)}={ A_{00}(\x)\over T_{0}},
\label{eq:nu5nu}
\end{align}
the chemical potential $\mu=e^{-\sigma}\mathbb{V}_{0}$ being associated with electric charge imbalance.
Our microscopic computation of the non-Abelian anomalous effective action leads to the following prediction for the 
couplings
\begin{align}
g_{1}(\nu)&=g_{2}(\nu)={N_{c}e^{2}\over 24\pi^{2}}\nu(\x) \nonumber \\[0.2cm]
g_{3}(\nu)&={N_{c} e^{2} \over 48\pi^2} \nu(\x)^{2} \label{eq:couplings_gs} \\[0.2cm]
g_{4}(\nu_{5})&={N_{c}e^{2}\over 48\pi^{2}}\nu_{5}(\x),
\nonumber
\end{align}
where the corresponding operators are gauge invariant. 
The equilibrium partition function \eqref{eq:WZW_2f_eff_EMfield} is the analog to 
Eq.~(2.26) of Ref.~\cite{Bhattacharyya:2012xi}, where the couplings \eqref{eq:couplings_gs} parametrize 
the action of a single NG boson associated with a broken U(1) symmetry. Similar formulae have been obtained in Ref.~\cite{Chapman:2013qpa}. 
Notice, however, that our expression is more general. In particular, it allows the computation of currents in sectors different from 
the electromagnetic one, as it will be shown later.

\section{Constitutive relations of the two-flavor hadronic superfluid}
\label{sec:currents}

The covariant currents computed above provide the first correction in the derivative expansion of the hydrodynamics constitutive relations
for the currents and the energy-momentum tensor. On general grounds, physical quantities in hydrodynamics admit the decomposition in terms
of their perfect fluid contributions and corrections containing higher terms in derivatives. For the case of a generic current
and the energy-momentum tensor, we have~\cite{Kovtun:2012rj,Bhattacharyya:2012xi,Chapman:2013qpa,Rezzolla:2013}
\begin{align}
\mathcal{J}^{\mu}&=\mathcal{J}^{\mu}_{\rm PF}+\nu^{\mu}, \nonumber \\[0.2cm]
T^{\mu\nu}&=T^{\mu\nu}_{\rm PF}+\Pi^{\mu\nu},
\label{eq:PF-NPF_gendecomp}
\end{align}
where the subscripts PF indicate the perfect fluid contributions and 
$\nu^{\mu}$ and $\Pi^{\mu\nu}$ are written in terms of the tensor quantities characterizing the system, as well as their spacetime derivatives. 
The 
constitutive relations provide the dependence of the coefficients of each term  
on physical quantities such as the temperature and the chemical potentials. Non-perfect fluids terms in~\eqref{eq:PF-NPF_gendecomp}
induce corrections in these constitutive relations at first and higher-order in derivatives. 

The presence of these gradient-dependent corrections to the perfect fluid quantities leads to ambiguities, 
since different definitions of the same physical variable may vary by gradient-dependent terms $(\delta T,\delta\mu,\delta u^{\mu},\ldots)$ 
which vanish in the limit of ideal hydrodynamics. Since these ambiguities cannot change the form of
the currents and the energy-momentum tensor, the perfect fluid constitutive relations have to compensate these
higher order terms by a shift in their perfect fluid values
\begin{align}
T_{0}&\longrightarrow T_{0}+\delta T(x),  \nonumber \\[0.2cm]
\mu_{0}&\longrightarrow \mu_{0}+\delta\mu(x), \\[0.2cm]
u^{\mu}_{0} &\longrightarrow u^{\mu}_{0}+\delta u^{\mu}(x).
\nonumber
\end{align} 
In hydrodynamics, these kind of ambiguities are fixed by selecting a frame, which in our case is chosen by requiring that one-derivative
corrections to perfect fluid quantities vanish. As a consequence, constributions at this order come only from the terms $\nu^{\mu}$ and
$\Pi^{\mu\nu}$ in Eq. \eqref{eq:PF-NPF_gendecomp}. For systems with spontaneous symmetry breaking, it was found~\cite{Manes:2018llx} 
that the energy-momentum tensor receives no corrections
\begin{align}
\Pi^{\mu\nu}&=0.
\end{align}
There are nevertheless nonvanishing corrections $\nu^{\mu}$ to the current, which admit the decomposition \eqref{eq:vector_KK-inv_decomp_pre} in terms of its longitudinal
and transverse components
\begin{align}
\nu^{\mu}=e^{\sigma}\big(\nu^{0}+a_{i}\nu^{i}\big)u^{\mu}+\big(\delta^{\mu}_{i}-\delta^{\mu}_{0}a_{i}\big)\nu^{i}.
\end{align} 
As explained in Sec. \ref{sec:anomaly}, the combinations $(\nu^{0}+a_{j}\nu^{j},\nu^{i})$ are KK-invariant.

\subsection{The gauge currents at leading order}

After these general considerations, we go back to the spontaneously broken two-flavor QCD model presented above,
and begin with the analysis of the gauge currents 
at leading order in the derivative expansion.
These and the energy-momentum tensor are computed by taking the corresponding 
functional derivatives on the effective action \eqref{eq:W_leading_orderSSB}. For the gauge currents, 
this calculation leads to the expressions\footnote{Since the BZ terms contain one 
derivative of the gauge fields (see Section \ref{subsec:currents_first_order}), at leading order in the derivative expansion there is no 
distinction between consistent and covariant currents. The subscript in the expectation value just indicates that they provide the
perfect fluid contribution.}
\begin{align}
\langle \mathcal{J}_{00}\rangle_{\rm PF} &\equiv -{T_0 e^{\sigma}\over \sqrt{g}} {\delta W \over \delta   V_{00}} = -n_{0}e^\sigma, \nonumber \\[0.2cm]
\langle \mathcal{J}_{30}\rangle_{\rm PF} &\equiv -{T_0 e^{\sigma}\over \sqrt{g}}{\delta W\over \delta   V_{30}} = -n_{3}e^\sigma
+{f_{\pi}^{2}\over 2}  V_{30} {\rm Tr\,}\Big([Q,U][Q,U^{\dagger}]\Big), \nonumber \\[0.2cm]
\langle \mathcal{J}^{i}_{0}\rangle_{\rm PF} &\equiv  {T_0 e^{-\sigma}\over \sqrt{g}} {\delta W\over \delta  V_{0i}} = 0, 
\label{eq:leading_Js}\\[0.2cm]
\langle \mathcal{J}^{i}_{3}\rangle_{\rm PF} &\equiv {T_0 e^{-\sigma}\over \sqrt{g}} {\delta W\over \delta  V_{3i}} = {if_\pi^{2}\over 4} 
g^{ij} \Tr \Big([Q,U] \Phi_{j}^{\dagger} + [Q,U^{\dagger}] \Phi_{j} \Big),
\nonumber
\end{align}
where the number densities $n_{0}$ and $n_{3}$ are defined by
\begin{align}
n_{0}={\partial P\over\partial \mu_{0}}, \hspace*{1cm}
n_{3}={\partial P\over\partial \mu_{3}}.
\end{align}
These densities satisfy the thermodynamic identity
\begin{align}
\varepsilon+P=Ts+\mu_{0}n_{0}+\mu_{3}n_{3},
\end{align}
where the entropy density is given by 
\begin{align}
s={\partial P\over \partial T}. 
\end{align}
In computing the currents \eqref{eq:leading_Js} we have also replaced $t_{3}$ by $Q$ inside the commutators. 

It is interesting to notice that, since the term $\mathscr{L}$ in \eqref{eq:W_leading_orderSSB} does not depend on $ V_{0\mu}$, it
does not contribute to the baryonic current. Then, the relations \eqref{eq:J_em,bar,iso_def} 
imply that the NG-dependent part of the consistent
electromagnetic current is completely determined by the isospin one. On the other hand, the
NG-independent term $P_0$ would contribute to both the baryonic and isospin currents, since
this function depends on both~$ V_{00}$ and $ V_{30}$.
Finally, the components of the energy-momentum tensor are similarly computed to give
\begin{align}
\langle T_{00}\rangle_{\rm PF}&\equiv -{T_0 e^{\sigma}\over \sqrt{g}}{\delta W\over \delta \sigma} 
=e^{2\sigma}\varepsilon+{f_\pi^2\over 2}{\rm Tr\,} \big(\xi_0 \xi_0^\dagger\big), \nonumber \\[0.2cm]
\langle T_{0}^{\,\,\,i}\rangle_{\rm PF} &\equiv {T_{0} e^{-\sigma}\over \sqrt{g}}\left({\delta W\over \delta a_{i}} 
- V_{00}{\delta W\over \delta  V_{0i}}- V_{30}{\delta W\over \delta  V_{3i}}\right)
={f_\pi^2\over 4}g^{ij}{\rm Tr\,}\Big(\xi_{0}\Phi_{j}^{\dagger}+\xi_{0}^{\dagger}\Phi_{j}\Big),
\label{eq:leading_emtensor} \\[0.2cm]
\langle T^{ij}\rangle_{\rm PF} &\equiv -{2T_{0}e^{\sigma}\over \sqrt{g}}g^{ik}g^{jm}{\delta W\over \delta g^{km}} 
=Pg^{ij}+{f_{\pi}^{2}\over 4}g^{ik}g^{jm}{\rm Tr\,}\Big(\Phi_{k}\Phi_{m}^{\dagger}+\Phi_{m}\Phi_{k}^{\dagger}\Big).
\nonumber
\end{align}
These expressions can be brought into a covariant form by writing them in terms of the static metric $G_{\mu\nu}$ defined in Eq.~\eqref{eq:static_line_element} and the 
four velocity $u_{\mu}$ given in~\eqref{eq:four_velocity}. For the currents, we find
\begin{align}
\langle \mathcal{J}_{0\mu}\rangle_{\rm PF}&=n_{0}u_{\mu}, \nonumber \\[0.2cm]
\langle \mathcal{J}_{3\mu}\rangle_{\rm PF}&=n_{3}u_{\mu}+{if_{\pi}^{2}\over 4}{\rm Tr\,}\Big([Q,U]\partial_{\mu}U^{\dagger}+
[Q,U^{\dagger}]\partial_{\mu}U\Big) \\[0.2cm]
&+{f_{\pi}^{2}\over 2} V_{3\mu}{\rm Tr\,}\Big([Q,U][Q,U^{\dagger}]\Big),
\nonumber
\end{align}
whereas the expression for the energy-momentum tensor is
\begin{align}
\langle T^{\mu\nu}\rangle_{\rm PF}&=(\epsilon+P)u^{\mu}u^{\nu}+PG^{\mu\nu}+{f_{\pi}^{2}\over 4}G^{\mu\alpha}G^{\nu\beta}
{\rm Tr\,}\Big[D_{\alpha}U(D_{\beta}U)^{\dagger}+D_{\beta}U(D_{\alpha}U)^{\dagger}\Big].
\end{align}

\subsection{Covariant currents and Bardeen-Zumino terms at first order in derivatives}
\label{subsec:currents_first_order}

The anomaly-induced WZW partition function \eqref{eq:general_SSB_U2xU2} provides the corrections at first order in derivatives
to the leading order currents and energy-momentum 
tensor given in Eqs.~\eqref{eq:leading_Js} and \eqref{eq:leading_emtensor}. 
It is at this order that we have to start distinguishing between consistent and covariant currents. The usual approach to compute them is to begin
with the consistent current, which is obtained by taking functional derivatives of the WZW partition function~\eqref{eq:general_SSB_U2xU2}. 
The covariant current is then found by adding the corresponding BZ terms. 
Here we use a more efficient procedure developed in \cite{Manes:2018llx} and reviewed in Section \ref{sec:anomaly}. 
Instead of using the WZW action, we arrive at the covariant current directly from the BZ current using Eqs. \eqref{eq:JR_cov_SSB_explicit} and 
\eqref{eq:JRJL_cov_SSB_explicit}.

The fact that the anomaly admits two different forms, consistent and covariant,
brings up the issue of which current is physically relevant for the evaluation of the anomalous transport coefficients. 
As far as the physics of transport is concerned, it is necessary to consider the covariant currents, as can be argued using the notion of 
anomaly inflow \cite{Callan:1984sa}. As seen in \cite{Jensen:2013kka,Haehl:2013hoa,Manes:2018llx}, 
the covariant anomaly is the result of the transport of conserved global charge in the higher-dimensional bulk space 
that overflows onto the spacetime boundary.
This strongly indicates that actual charge transport is governed by the covariant current,
while the consistent current determine the coupling to the sources in the equilibrium partition function\footnote{The physical relevance of the covariant
current in a different context has been discussed in \cite{Iso:2006wa}.}. 

The upshot is that covariant currents, although not coupling directly to the external fields, 
are the ones relevant in the hydrodynamic description and provide the constitutive relations
that lead to the determination of the different transport coefficients. In the case at hand, 
we particularize Eq.~\eqref{eq:JR_cov_SSB_explicit} to the background defined in 
\eqref{eq:2fQCD_background_t0t3}, keeping always in mind that $ A_{00}=e^{\sigma}\mu_{5}$ is a constant field. 
Then, we apply the relations \eqref{eq:JRJL_cov_SSB_explicit} leading to the calculation of the vector and axial-vector
components of the current. After some algebra, we arrive at 
\begin{align}
\langle \mathcal{J}^{\mu}_{0V}\rangle_{\rm cov}&=-{N_{c}\over 16\pi^{2}}\epsilon^{\mu\nu\alpha\beta}\bigg\{\mcA_{0\nu}\mcV_{0\alpha\beta}
-{1\over 2}\mcV_{3\nu\alpha}{\rm Tr\,}\Big[\Big(R_{\beta}+L_{\beta}\Big)Q\Big]
\nonumber \\[0.2cm]
&\left.+\mcV_{3\nu}\partial_{\alpha}{\rm Tr\,}\Big[\Big(R_{\beta}+L_{\beta}\Big)Q\Big]
-{i\over 3}{\rm Tr\,}\Big(L_{\nu}L_{\alpha}L_{\beta}\Big)
\right\}, 
\nonumber  \\[0.2cm]
\langle \mathcal{J}^{\mu}_{3V}\rangle_{\rm cov}&=-{N_{c}\over 48\pi^{2}}\epsilon^{\mu\nu\alpha\beta}\bigg(
3\mcA_{0\nu}\mcV_{3\alpha\beta}
+\mcA_{0\nu}\mcV_{3\alpha\beta}{\rm Tr\,}\Big[\Big(U^{-1}QU-Q\Big)Q\Big]
\label{eq:JcovaVcomponents_cov} \\[0.2cm]
&-\partial_{\alpha}\Big\{\mcA_{0\nu}{\rm Tr\,}\Big[\big(R_{\beta}-L_{\beta}\big)Q\Big]
+2\mcA_{0\nu}\mcV_{3\beta}{\rm Tr\,}\Big[\Big(U^{-1}QU-Q\Big)Q\Big]
\Big\} \nonumber \\[0.2cm]
&-{3\over 2}\mcV_{0\nu\alpha}{\rm Tr\,}\Big[\big(R_{\beta}+L_{\beta})Q\Big]
\bigg),
\nonumber 
\end{align}
for the vector currents. In the case of the axial-vector currents, a similar calculation leads to the results 
\begin{align}
\langle \mathcal{J}^{\mu}_{0A}\rangle_{\rm cov}&=-{N_{c}\over 48\pi^{2}}\epsilon^{\mu\nu\alpha\beta}\bigg(\mcA_{0\nu}\mcA_{0\alpha\beta}
-{1\over 2}\mcV_{3\nu\alpha}\Big\{{\rm Tr\,}\Big[\big(R_{\beta}-L_{\beta}\big)Q\Big]
\nonumber \\[0.2cm]
&+2\mcV_{3\beta}{\rm Tr\,}\Big[\Big(U^{-1}QU-Q\Big)Q\Big]\Big\}
\bigg),
\label{eq:JcovaAcomponents_cov} \\[0.2cm]
\langle \mathcal{J}^{\mu}_{3A}\rangle_{\rm cov}&=-{N_{c}\over 48\pi^{2}}\epsilon^{\mu\nu\alpha\beta}\left(
-{3\over 2}\mcV_{0\nu\alpha}\left\{{\rm Tr\,}\Big[\big(R_{\beta}-L_{\beta})t_{3}\Big]
+2\mcV_{3\beta}{\rm Tr\,}\Big[\Big(U^{-1}QU-Q\Big)Q\Big]\right\}
\right.\nonumber \\[0.2cm]
&-{1\over 2}\mcA_{0\nu\alpha}{\rm Tr\,}\Big[\big(R_{\beta}+L_{\beta}\big)Q\Big]
+\mcA_{0\nu}\partial_{\alpha}{\rm Tr\,}\Big[\Big(R_{\beta}+L_{\beta}\Big)Q\Big]\bigg).
\nonumber
\end{align} 
We will make use of these covariant expressions in later sections to identify the relevant tensor structures and 
compute the out-of-equilibrium transport coefficients. 

In order to find the consistent currents, we need to find the explicit form of the BZ terms \eqref{eq:BZ_current_VA_gen}
in the background \eqref{eq:2fQCD_background_t0t3}.
For the vector BZ currents we have ($a=0,3$)
\begin{align}
\langle \mathcal{J}^{\mu}_{aV}\rangle_{\rm BZ}&=-{N_{c}\over 4\pi^{2}}\epsilon^{\mu\nu\alpha\beta}{\rm Tr\,}\Big(t_{a}\mcV_{\nu\alpha}\mcA_{\beta}\Big)
=-{N_{c}\over 16\pi^{2}}\epsilon^{\mu\nu\alpha\beta}\mcA_{0\nu}\mcV_{a\alpha\beta}.
\label{eq:JVBZ_scriptfonts}
\end{align}
Using Eq. \eqref{eq:cov_form_mbhA_mu}, it can be written as
\begin{align}
\langle\mathcal{J}^{\mu}_{aV}\rangle_{\rm BZ}&={N_{c}\over 8\pi^{2}}e^{-\sigma} A_{00}\mcB^{\mu}_{a}
={N_{c}\over 8\pi^{2}}\mu_{5}\mcB^{\mu}_{a},
\label{eq:JVBZ_pre}
\end{align}
where the magnetic field \eqref{eq:magnetic_field} has been identified.

At this point, it is important to stress that
since in the background \eqref{eq:2fQCD_background_t0t3}
all fields lie on the Cartan subalgebra of the group the theory is Abelianized, so the term cubic in $\mcA_{\mu}$ in the
general expression of the BZ current 
vanishes. We should remember, however, that this formally vanishing cubic term has to be kept when using the BZ current 
to compute the covariant currents
applying  \eqref{eq:JRJL_cov_SSB}. The reason is that the term gives a nonvanishing contribution
upon transforming the left gauge field, as it is required by our prescription. Our construction of the covariant currents 
presented above used the explicit expression \eqref{eq:JR_cov_SSB_explicit}, which already includes this contribution.

A similar analysis can be carried out for the axial-vector BZ current shown in \eqref{eq:BZ_current_VA_gen}, 
which in the background of interest takes the form
\begin{align}
\langle\mathcal{J}^{\mu}_{aA}\rangle_{\rm BZ}&=-{N_{c}\over 12\pi^{2}}\epsilon^{\mu\nu\alpha\beta}{\rm Tr\,}\Big(t_{a}\mcA_{\nu}\mcA_{\alpha\beta}\Big)
=-{N_{c}\delta_{a0}\over 24\pi^{2}}\epsilon^{\mu\nu\alpha\beta}\mcA_{0\nu}\partial_{\alpha}\mcA_{0\beta}
\nonumber \\[0.2cm]
&=-{N_{c}\delta_{a0}\over 12\pi^{2}}e^{-2\sigma} A_{00}^{2}\omega^{\mu}
=-{N_{c}\delta_{a0}\over 12\pi^{2}}\mu_{5}^{2}\omega^{\mu},
\label{eq:JABZ_scriptfonts}
\end{align}
where we used the vorticity vector \eqref{eq:vorticity_field} and introduced the Kronecker delta to indicate that 
$\langle\mathcal{J}^{\mu}_{3A}\rangle_{\rm BZ}$ vanishes identically 
\begin{align}
\langle\mathcal{J}^{\mu}_{3A}\rangle_{\rm BZ}=0.
\label{eq:JBZ3A=0}
\end{align}

Once the BZ currents are known, we compute the consistent gauge currents using 
\begin{align}
\langle \mathcal{J}_{aV,A}^{\mu}\rangle_{\rm cons}=\langle \mathcal{J}_{aV,A}^{\mu}\rangle_{\rm cov}
-\langle \mathcal{J}_{aV,A}^{\mu}\rangle_{\rm BZ}.
\end{align}
Subtracting the BZ terms computed in Eqs. \eqref{eq:JVBZ_scriptfonts} and \eqref{eq:JABZ_scriptfonts} from the covariant currents
\eqref{eq:JcovaVcomponents_cov} and \eqref{eq:JcovaAcomponents_cov}, we find that the consistent currents can be identified 
from the covariant ones just by dropping all terms independent of the NG boson matrix $U$. 
The consistent currents can be alternatively computed
taking functional derivatives from the WZW effective action \cite{Banerjee:2012iz}
\begin{align}
\langle \mathcal{J}_{a0V}\rangle_{\rm cons}&=-{T_{0}e^{\sigma}\over \sqrt{g}}{\delta W\over \delta V_{a0}}, \nonumber \\[0.2cm]
\langle \mathcal{J}_{aV}^{i}\rangle_{\rm cons}&={T_{0}e^{-\sigma}\over \sqrt{g}}{\delta W\over \delta V_{ai}}, \nonumber \\[0.2cm]
\langle \mathcal{J}_{a0A}\rangle_{\rm cons}&=-{T_{0}e^{\sigma}\over \sqrt{g}}{\delta W\over \delta A_{a0}}, \\[0.2cm]
\langle \mathcal{J}_{aA}^{i}\rangle_{\rm cons}&={T_{0}e^{-\sigma}\over \sqrt{g}}{\delta W\over \delta A_{ai}}.
\nonumber
\end{align}
However, in computing the spatial components of the axial-vector currents, whose dual gauge field are set to zero in the background
\eqref{eq:2fQCD_background_t0t3}, we have to carry out the functional derivatives in the WZW action before actually setting these fields to zero, 
which notably complicates the calculation.
The procedure used here avoids this,
since it only 
makes use of the BZ currents, being much more economical from a computational point of view.

\subsection{Corrections to the leading order constitutive relations}
\label{sec:corrections_LO}

A look at the results for the vector and axial-vector covariant currents given 
in Eqs. \eqref{eq:JcovaVcomponents_cov} and \eqref{eq:JcovaAcomponents_cov} shows that all dependence on the NG bosons matrix $U$ comes in terms of
the following covariant expressions
\begin{align}
\mathds{H}&\equiv {\rm Tr\,}\Big[\Big(U^{-1}QU-Q\Big)Q\Big], \nonumber \\[0.2cm]
\mathds{I}_{\mu}&\equiv {\rm Tr\,}\Big[\Big(R_{\mu}+L_{\mu}\Big)Q\Big], \\[0.2cm]
\mathds{T}_{\mu}&\equiv {\rm Tr\,}\Big[Q\Big(R_{\mu}-L_{\mu}\Big)\Big]+2\mcV_{3\mu}{\rm Tr\,}\Big[\Big(U^{-1}QU-Q\Big)Q\Big].
\nonumber
\end{align}
In terms of them, the one-derivative corrections to the currents in the static background~\eqref{eq:static_line_element}
take the form
\begin{align}
\langle\mathcal{J}^{\mu}_{0V}\rangle_{\rm cov}&=
{N_{c}\over 16\pi^{2}}\epsilon^{\mu\nu\alpha\beta}
\left[\partial_{\nu}\mcV_{3\alpha}\mathds{I}_{\beta}-\mcV_{3\nu}\partial_{\alpha}\mathds{I}_{\beta}
+{i\over 3}{\rm Tr\,}\Big(L_{\nu}L_{\alpha}L_{\beta}\Big)\right]
+{N_{c}\over 8\pi^{2}}\mu_{5}\mcB_{0}^{\mu}, 
\nonumber \\[0.2cm]
\langle\mathcal{J}^{\mu}_{3V}\rangle_{\rm cov}&=-{N_{c}\over 48\pi^{2}}\epsilon^{\mu\nu\alpha\beta}
\bigg[\partial_{\alpha}\Big(\mu_{5}u_{\nu}\mathds{T}_{\beta}\Big)
-3\partial_{\nu}\mcV_{0\alpha}\mathds{I}_{\beta}\bigg]+
{N_{c}\over 24\pi^{2}}\mu_{5}(\mathds{H}+3)\mcB^{\mu}_{3},
\end{align}
and
\begin{align}
\langle\mathcal{J}^{\mu}_{0A}\rangle_{\rm cov}&=
{N_{c}\over 48\pi^{2}}\epsilon^{\mu\nu\alpha\beta}\Big(
\partial_{\nu}\mcV_{3\alpha}\mathds{I}_{\beta}+2\mathds{H}\partial_{\nu}\mcV_{3\alpha}\mcV_{3\beta}\Big)
-{N_{c}\over 12\pi^{2}}
\mu_{5}^{2}\omega^{\mu}, 
\nonumber \\[0.2cm]
\langle\mathcal{J}^{\mu}_{3A}\rangle_{\rm cov}&={N_{c}\over 48\pi^{2}}\epsilon^{\mu\nu\alpha\beta}
\Big[3\partial_{\nu}\mcV_{0\alpha}\mathds{T}_{\beta}-\partial_{\nu}\big(\mu_{5}u_{\alpha}\big)\mathds{I}_{\beta}
+\mu_{5}u_{\nu}\partial_{\alpha}\mathds{I}_{\beta}\Big].
\end{align}

The currents can now be decomposed into their longitudinal and transverse components with respect to the four-velocity $u_{\mu}$. 
The results can be written as linear combinations of the following five pseudo-scalar quantities
\begin{align}
\mathds{S}_{1,a}&\equiv \epsilon^{\mu\nu\alpha\beta}\mathds{I}_{\mu}u_{\nu}\partial_{\alpha}\mcV_{a\beta}
=\mathds{I}_{\mu}\mcB^{\mu}_{a} \hspace*{2cm}
(a=0,3), \nonumber \\[0.2cm]
\mathds{S}_{2}&\equiv {1\over 2}\epsilon^{\mu\nu\alpha\beta}\mathds{I}_{\mu}u_{\nu}\partial_{\alpha}u_{\beta}
=\mathds{I}_{\mu}\omega^{\mu}, \nonumber \\[0.2cm]
\mathds{S}_{3}&\equiv \epsilon^{\mu\nu\alpha\beta}u_{\mu}\left[\mcV_{3\nu}\partial_{\alpha}\mathds{I}_{\beta}
-{i\over 3}{\rm Tr\,}\Big(L_{\nu}L_{\alpha}L_{\beta}\Big)\right], 
\label{eq:pseudoscalarsSs}\\[0.2cm]
\mathds{S}_{4,a}&\equiv \epsilon^{\mu\nu\alpha\beta}\mathds{T}_{\mu}u_{\nu}\partial_{\alpha}\mcV_{a\beta}=\mathds{T}_{\mu}\mcB_{a}^{\mu}
\hspace*{2cm} (a=0,3),  \nonumber \\[0.2cm]
\mathds{S}_{5}&\equiv {1\over 2}\epsilon^{\mu\nu\alpha\beta}\mathds{T}_{\mu}u_{\nu}\partial_{\alpha}u_{\beta}=\mathds{T}_{\mu}\omega^{\mu}, \nonumber
\end{align}
together with the four transverse pseudo-vectors
\begin{align}
\mathds{P}^{\mu}_{1,a}&\equiv \epsilon^{\mu\nu\alpha\beta}u_{\nu}\mathds{I}_{\alpha}\partial_{\beta}\left({\mu_{a}\over T}\right) \hspace*{2cm}
(a=0,3),
\nonumber \\[0.2cm]
\mathds{P}^{\mu}_{2}&\equiv \epsilon^{\mu\nu\alpha\beta}u_{\nu}\partial_{\alpha}\mathds{I}_{\beta}, \nonumber \\[0.2cm]
\mathds{P}^{\mu}_{3,a}&\equiv \epsilon^{\mu\nu\alpha\beta}u_{\nu}\mathds{T}_{\alpha}\partial_{\beta}\left({\mu_{a}\over T}\right) \hspace*{2cm}
(a=0,3), \label{eq:pseudovectorPs}\\[0.2cm]
\mathds{P}^{\mu}_{4}&\equiv \epsilon^{\mu\nu\alpha\beta}u_{\nu}\partial_{\alpha}\mathds{T}_{\beta}.
\nonumber 
\end{align} 
Notice that the definitions in Eq. \eqref{eq:pseudoscalarsSs} involve either the 
magnetic field or the vorticity vector, defined in Eqs. \eqref{eq:magnetic_field} and
\eqref{eq:vorticity_field} respectively. 
The one-derivative corrections to the gauge currents can now be written as 
\begin{align}
u_{\mu}\langle \mcJ^{\mu}_{0V}\rangle_{\rm cov}&=-{N_{c}\over 16\pi^{2}}\Big(\mathds{S}_{1,3}+\mathds{S}_{3}\Big),
\nonumber \\[0.2cm]
P^{\mu}_{\,\,\,\nu}\langle \mcJ^{\nu}_{0V}\rangle_{\rm cov}&=-{N_{c}\over 16\pi^{2}}
\Big(T\mathds{P}^{\mu}_{1,3}
-\mu_{3}\mathds{P}^{\mu}_{2}
-2\mu_{5}\mcB^{\mu}_{0}\Big),
\label{eq:vector_currents_cov_lontrans1}
\end{align}
for the zero flavor components, while the three-component gives 
\begin{align}
u_{\mu}\langle \mcJ^{\mu}_{3V}\rangle_{\rm cov}&=-{N_{c}\over 48\pi^{2}}\Big(
3\mathds{S}_{1,0}
+2\mu_{5}\mathds{S}_{5}\Big), 
\nonumber \\[0.2cm]
P^{\mu}_{\,\,\,\nu}\langle \mcJ^{\nu}_{3V}\rangle_{\rm cov}&=-{N_{c}\over 48\pi^{2}}\Big[
3T\mathds{P}_{1,0}^{\mu}
+\mu_{5}\mathds{P}_{4}^{\mu}+4\mu_{3}\mu_{5}\mathds{H}\omega^{\mu}-2\mu_{5}(\mathds{H}+3)\mcB^{\mu}_{3}\Big].
\label{eq:vector_currents_cov_lontrans2}
\end{align}
In all these equations, we have used that fields are time-independent (and in particular 
$ A_{00}=-\mu_{5}u_{0}$ is constant), as well as the expression of the local temperature 
$T(\x)=T_{0}e^{-\sigma(\x)}$. In addition, we applied the identity
\begin{align}
P^{\mu}_{\,\,\,\nu}\epsilon^{\nu\omega\alpha\beta}\mathcal{T}_{\omega\alpha\beta}
=-\epsilon^{\mu\nu\alpha\beta}u_{\nu}e^{-\sigma}\Big(\mathcal{T}_{0\alpha\beta}-\mathcal{T}_{\beta\alpha0}\Big),
\label{eq:useful_id_proj2}
\end{align}
valid for any tensor $\mathcal{T}_{\mu\alpha\beta}$ transverse in the $\alpha$ index, 
$\mathcal{T}_{\mu\alpha\beta}=\mathcal{T}_{\mu\nu\beta}P^{\nu}_{\,\,\,\alpha}$. This equation can be applied
when the last index in $\mathcal{T}_{\mu\alpha\beta}$ is
also transverse, $\mathcal{T}_{\mu\alpha\beta}=\mathcal{T}_{\mu\nu\sigma}P^{\nu}_{\,\,\,\alpha}P^{\sigma}_{\,\,\,\beta}$, 
in which case the second term on the right-hand side of \eqref{eq:useful_id_proj2} is equal to zero.

We see how the covariant vector currents can be written solely in terms of 
the pseudoscalar structures $\mathds{S}_{1,a}$, $\mathds{S}_{3}$, and $\mathds{S}_{5}$, as well as
the pseudovectors $\mathds{P}^{\mu}_{1,a}$, $\mathds{P}^{\mu}_{2}$, and $\mathds{P}^{\mu}_{4}$. The 
remaining structures, which will appear in the vector-axial currents, have zero transport coefficients
associated. The contribution of the BZ terms to Eqs. \eqref{eq:vector_currents_cov_lontrans1} and
\eqref{eq:vector_currents_cov_lontrans2} can be easily identified as those terms proportional to 
the magnetic field $\mcB^{\mu}_{a}$ without gothic fonts prefactors [cf. \eqref{eq:JVBZ_pre}]. Of course, these
terms can only appear in the transverse components of the covariant current, the BZ terms being themselves transverse.  
Notice as well that the expressions of the covariant currents are given only in terms of 
the KK-variant gauge fields $\mcV_{a\mu}$ and $\mcA_{0\mu}$, without any explicit reference
to the KK gauge field $a_{i}$. 

Introducing the electric fields for $a=0,3$
\begin{align}
\mcE_{a\mu}=\mcV_{a\mu\nu}u^{\nu}=e^{-\sigma}\partial_{\mu}\mcV_{a0}=T\partial_{\mu}\left({\mu_{a}\over T}\right),
\end{align}
we can write the first pseudovector in \eqref{eq:pseudovectorPs} as
\begin{align}
\mathds{P}^{\mu}_{1,a}&={1\over T}\epsilon^{\mu\nu\alpha\beta}u_{\nu}\mathds{I}_{\alpha}\mcE_{a\beta},
\end{align}
and identify the terms proportional to $\mathds{P}^{\mu}_{1,0}$ and $\mathds{P}^{\mu}_{1,3}$ in the constitutive relations 
as the ones associated to the 
chiral electric effect~\cite{Neiman:2011mj}. In particular, we can construct the transverse covariant electromagnetic current as
\begin{align}
P^{\mu}_{\,\,\,\nu}\langle \mcJ^{\nu}_{\rm em}\rangle_{\rm cov}&=-{eN_{c}\over 48\pi^{2}}
\Big[\epsilon^{\mu\nu\alpha\beta}u_{\nu}\mathds{I}_{\alpha}\Big(3\mcE_{0\beta}+\mcE_{3\beta}\Big)
-\mu_{3}\epsilon^{\mu\nu\alpha\beta}u_{\nu}\partial_{\alpha}\mathds{I}_{\beta}
+\mu_{5}\epsilon^{\mu\nu\alpha\beta}u_{\nu}\partial_{\alpha}\mathds{T}_{\beta} \nonumber \\[0.2cm]
&+4\mu_{3}\mu_{5}\mathds{H}\omega^{\mu}-2\mu_{5}\mathds{H}\mcB^{\mu}_{3}-2\mu_{5}\Big(\mcB^{\mu}_{0}+3\mcB^{\mu}_{3}\Big)\Big].
\label{eq:emtrasnverse_current}
\end{align}
Using \eqref{eq:cov_electric_field_vs_E3E0} and \eqref{eq:GB_expansion_pions}, 
this can be written in terms of the electric $\mcE_{\mu}$, magnetic $\mcB_{\mu}$ and pions fields as
\begin{align}
P^{\mu}_{\,\,\,\nu}\langle \mcJ^{\nu}_{\rm em}\rangle_{\rm cov}&=
{e^{2}N_{c}\over 12\pi^{2}f_{\pi}}\epsilon^{\mu\nu\alpha\beta}u_{\nu}\partial_{\alpha}\pi^{0}\mcE_{\beta}
-{i\mu_{5}eN_{c}\over 12\pi^{2} f_{\pi}^{2}}\epsilon^{\mu\nu\alpha\beta}u_{\nu}\partial_{\alpha}\pi^{+}\partial_{\beta}\pi^{-} \nonumber \\[0.2cm]
&+{\mu_{5}e^{2}N_{c}\over 12\pi^{2} f_{\pi}^{2}}\epsilon^{\mu\nu\alpha\beta}u_{\nu}\partial_{\alpha}\big(\pi^{+}\pi^{-}\big)
\mcV_{\beta}
+{\mu_{3}\mu_{5}eN_{c}\over 6\pi^{2}f_{\pi}^{2}}\pi^{+}\pi^{-}\omega^{\mu} 
\label{eq:emtrasnverse_current_EB_fields}\\[0.2cm]
&+{5\mu_{5}e^{2}N_{c}\over 36\pi^{2}}\mcB^{\mu}+\mathcal{O}(\pi^{3}).
\nonumber
\end{align}
Interestingly, the term proportional to $\pi^{+}\pi^{-}\mcB^{\mu}$ cancels out from this expression. 

The emergence of the chiral electric effect, i.e. charge transport normal to the direction of the electric field, is manifest from the first term on 
Eq. \eqref{eq:emtrasnverse_current_EB_fields}, with a 
transport coefficient that can be identified from this equation. Despite the ongoing discussion in the literature concerning its nondissipative
character (see, for example, Ref.~\cite{Chapman:2013qpa}), our derivation of this term clearly 
shows that the chiral electric effect is intrinsically nondissipative, i.e. it does not lead to the production of entropy. Similar 
terms are present in the baryon and isospin currents.

In fact, the fundamental feature preserving the nondissipative character of the chiral electric effect in the constitutive relation \eqref{eq:emtrasnverse_current_EB_fields} is the appearance  of the pion field having negative signature under time reversal $\mathsf{T}$.
Indeed, given the transformation properties of the pions, $\mathsf{T}:\pi^{0}\rightarrow -\pi^{0}$, $\mathsf{T}:\pi^{\pm}\rightarrow -\pi^{\mp}$, the 
corresponding transport coefficient must be $\mathsf{T}$-even, thus revealing the reversible nature of the phenomenon. The situation is somewhat reminiscent of
the quantum Hall effect, where nondissipative electric charge transport appears triggered by the presence of the $\mathsf{T}$-odd external magnetic field.
The case of the chiral hadronic fluid studied here, however, has a closer resemblance to the {\em anomalous} quantum Hall effect, where nondissipative transport is the result
of $\mathsf{T}$-breaking brought about by the nontrivial topology of the band~\cite{Haldane:1988zza}.

To compare our results with other analysis in the literature, we use 
the identities \eqref{eq:GB_expansion_pions} to write the spatial components of this current in terms of the pion fields
and the electric and magnetic fields associated with the electromagnetic potential $\mathbb{V}_{\mu}$
as
\begin{align}
\langle \mcJ^{i}_{\rm em}\rangle_{\rm cov}&=
{e^{2}N_{c}\over 12\pi^{2}f_{\pi}}\epsilon^{ijk}\partial_{j}\pi^{0}\mathbb{E}_{k}
-{i\mu_{5}eN_{c}\over 12\pi^{2} f_{\pi}^{2}}\epsilon^{ijk}\partial_{j}\pi^{+}\partial_{k}\pi^{-} \nonumber \\[0.2cm]
&+{\mu_{5}e^{2}N_{c}\over 12\pi^{2} f_{\pi}^{2}}\epsilon^{ijk}\partial_{j}\big(\pi^{+}\pi^{-}\big)
\mathbb{V}_{k}
+{\mu\mu_{5}eN_{c}\over 6\pi^{2}f_{\pi}^{2}}\pi^{+}\pi^{-}\omega^{i}
\label{eq:emtrasnverse_current_pions} \\[0.2cm]
&+{5\mu_{5}e^{2}N_{c}\over 36\pi^{2}}\big(\mathbb{B}^{i}-2\mu\omega^{i}\big)+\mathcal{O}(\pi^{3}),
\nonumber
\end{align}
where we have introduced the components of the physical electric field $\mathbb{E}_{i}=e^{-\sigma}\partial_{i}\mathbb{V}_{0}$. 
In addition, we have introduced the 
electric charge chemical potential $\mu\equiv e^{-\sigma}\mathbb{V}_{0}$ [cf. Eqs. \eqref{eq:relationmu0mu3V0} and \eqref{eq:nu5nu}].
This covariant electromagnetic
current can be alternatively found by adding the corresponding BZ term to
the consistent current obtained by varying the 
WZW equilibrium partition function [cf. \eqref{eq:pion_eff_action_V0}] 
\begin{align}
W[\mathbb{V}_{\mu},\pi^{0},\pi^{\pm}]_{\rm WZW}
&=-\int\limits_{S^{3}}d^{3}x\sqrt{g}\, \left[{ e^{2}N_{c}\over 12\pi^{2} f_{\pi}T}\mu
\partial_{k}\pi^{0}\left(\mathbb{B}^{k}-\mu\omega^{k}\right)
\right.
\label{eq:pion_eff_action_mu3}\\[0.2cm]
&+\left.{i\mu_{5}eN_{c}\over 24\pi^{2} f_{\pi}^{2}T}\Big(
\pi^{-}\partial_{k}\pi^{+}-\pi^{+}\partial_{k}\pi^{-}-2ie\pi^{+}\pi^{-}\mathbb{V}_{k}\Big)
\Big(\mathbb{B}^{k}-2\mu \omega^{k}\Big)\right],
\nonumber
\end{align}
with respect to the electromagnetic vector potential $\mathbb{V}_{i}$. 
Similar expressions can be written for the baryon and isospin currents. In the first case, we find
\begin{align}
\langle \mcJ^{i}_{\rm bar}\rangle_{\rm cov}&={eN_{c}\over 12\pi^{2}}\left[{1\over f_{\pi}}\epsilon^{ijk}\partial_{j}\pi^{0}
\mathbb{E}_{k}+{1\over 3}\mu_{5}\Big(\mathbb{B}^{i}-2\mu\omega^{i}\Big)\right]+\mathcal{O}(\pi^{3}),
\label{eq:baryon_current_pions}
\end{align}
while for the isospin current the result is
\begin{align}
\langle \mcJ^{i}_{\rm iso}\rangle_{\rm cov}&=-{eN_{c}\over 24\pi^{2}}\left[
-{1\over f_{\pi}}\epsilon^{ijk}\partial_{j}\pi^{0}\mathbb{E}_{k}
+{2i\over ef_{\pi}^{2}}\mu_{5}\epsilon^{ijk}\partial_{j}\pi^{+}\partial_{k}\pi^{-}
-{2\over f_{\pi}^{2}}\mu_{5}\epsilon^{ijk}\partial_{j}\big(\pi^{+}\pi^{-}\big)\mathbb{V}_{k} \right.\nonumber \\[0.2cm]
&\left.-{4\over f_{\pi}^{2}}\mu\mu_{5}\pi^{+}\pi^{-}\omega^{i}
-3\mu_{5}\Big(\mathbb{B}^{i}-2\mu\omega^{i}\Big)\right]+\mathcal{O}(\pi^{3}).
\label{eq:isospin_current_pions}
\end{align}
 All three currents are easily seen to be invariant under the gauge transformations of electromagnetism 
$\delta_{\varepsilon}\mathbb{V}_{i}=\partial_{i}\varepsilon$, $\delta_{\varepsilon}\pi^{\pm}=\pm ie\varepsilon\pi^{\pm}$.

A relevant feature to notice in all three currents shown in 
Eqs. \eqref{eq:emtrasnverse_current_pions}, \eqref{eq:baryon_current_pions}, 
and \eqref{eq:isospin_current_pions} is that
the terms proportional to the vorticity vector $\omega^{i}$ come always 
multiplied by the chiral chemical potential $\mu_{5}$. This means that
in the absence of chiral imbalance ($\mu_{5}=0$) there are no contributions to the electromagnetic, isospin, and baryonic currents 
depending of the vorticity. 

Despite of this fact, the partition function \eqref{eq:pion_eff_action_mu3} shows the existence of a direct coupling of the 
vorticity vector to the pion gradient which is not weighted by $\mu_{5}$. Setting the electromagnetic field to zero for simplicity, this leads to the following form of
the isospin number density 
\begin{align}
n_{3}={({\rm Tr\,}Q)eN_{c}\over 2\pi^{2}f_{\pi}}\mu\partial_{k}\pi^{0}\omega^{k},
\end{align} 
where we have written the result in terms of the trace of the charge matrix $Q$. This expression 
agrees with the results found
in~\cite{Huang:2017pqe} for the baryon and isospin number densities of a hadronic fluid under rotation \cite{Fukushima:2018grm}, 
which shows the existence of a 
vorticity coupling with the pion gradient
in the absence of chiral imbalance. It is nevertheless surprising that no such effect survives in the spatial currents themselves, as it is seen 
from our explicit results.  

To close this section, we study the one-derivative corrections to the constitutive relations derived from the covariant axial-vector currents. Their longitudinal
and transverse projections are now given by
\begin{align}
u_{\mu}\langle \mcJ^{\mu}_{0A}\rangle_{\rm cov}&=-{N_{c}\over 48\pi^{2}}\mathds{S}_{4,3},
\nonumber \\[0.2cm]
P^{\mu}_{\,\,\,\nu}\langle \mcJ^{\nu}_{0A}\rangle_{\rm cov}&=-{N_{c}\over 48\pi^{2}}\Big(T\mathds{P}^{\mu}_{3,3}
+2\mu_{3}\mathds{H}\mcB_{3}^{\mu}+4\mu_{5}^{2}\omega^{\mu}\Big),
\label{eq:axial_currents_cov_lontrans1}
\end{align}
and
\begin{align}
u_{\mu}\langle \mcJ^{\mu}_{3A}\rangle_{\rm cov}&=-{N_{c}\over 48\pi^{2}}\Big(
3\mathds{S}_{4,0}
-2\mu_{5}\mathds{S}_{2}\Big), 
\nonumber \\[0.2cm]
P^{\mu}_{\,\,\,\nu}\langle \mcJ^{\nu}_{3A}\rangle_{\rm cov}&=-{N_{c}\over 48\pi^{2}}
\Big(3T\mathds{P}^{\mu}_{3,0}+6\mu_{3}\mathds{H}\mcB_{0}^{\mu}-
\mu_{5}\mathds{P}^{\mu}_{2}\Big).
\label{eq:axial_currents_cov_lontrans2}
\end{align}
As in the case of the vector currents, here we also find the occurrence of a nondissipative chiral electric effect associated with the 
terms $\mathds{P}^{\mu}_{3,a}$, whose corresponding susceptibilities are obtained from the previous equations.  

Expressed in terms of the KK-invariant electromagnetic fields, we find the following explicit expressions for the transverse components of the
axial-vector covariant currents
\begin{align}
\langle \mathcal{J}^{i}_{0A}\rangle_{\rm cov}&={N_{c}\over 24\pi^{2}f_{\pi}^{2}}\Big[
ie\epsilon^{ijk}\Big(\pi^{-}\partial_{j}\pi^{+}-\pi^{+}\partial_{j}\pi^{-}-
2ie\pi^{+}\pi^{-}\mathbb{V}_{j}\Big)\mathbb{E}_{k} \nonumber\\[0.2cm]
&+2e\mu\pi^{+}\pi^{-}\Big(\mathbb{B}^{i}-2\mu\omega^{i}\Big)-2\mu_{5}^{2}\omega^{i}\Big]+\mathcal{O}(\pi^{3}), \nonumber \\[0.2cm]
\langle \mathcal{J}^{i}_{3A}\rangle_{\rm cov}&={N_{c}\over 24\pi^{2}f_{\pi}^{2}}\Big[
ie\epsilon^{ijk}\Big(\pi^{-}\partial_{j}\pi^{+}-\pi^{+}\partial_{j}\pi^{-}
-2ie\pi^{+}\pi^{-}\mathbb{V}_{j}\Big)\mathbb{E}_{k}
\label{eq:axial_KK_inv_comps}\\[0.2cm]
&-2e\pi^{+}\pi^{-}\Big(\mathbb{B}^{i}-2\mu\omega^{i}\Big)\Big]+\mathcal{O}(\pi^{3}).
\nonumber
\end{align}
These constitutive relations  
contain chiral separation effect terms of electric, magnetic, and vortical type. 
Although all explicit dependence on the
vorticity in Eq.~\eqref{eq:axial_currents_cov_lontrans1} comes from the BZ term and seems to disappear in the absence of chiral imbalance ($\mu_{5}=0$),
once written in terms of the KK-invariant quantities we find 
vorticity-dependent terms in both constitutive relations, 
mediated by the charged pion fields, that survive in this limit.

Our calculation based on Eq. \eqref{eq:JRJL_cov_SSB} shows clearly that in the limit of vanishing NG fields, $\zeta_{a}\rightarrow 0$ ($U\rightarrow \mathbb{1}$),
the covariant currents are given by the BZ terms. For the electromagnetic transverse current, we find from Eq.
\eqref{eq:emtrasnverse_current_pions}
\begin{align}
\lim_{\pi^{0},\pi^{\pm}\rightarrow 0}P^{\mu}_{\,\,\,\nu}\langle \mcJ^{\nu}_{\rm em}\rangle_{\rm cov}={5\mu_{5}e^{2}N_{c}\over 36\pi^{2}}\mcB^{\mu},
\end{align}
where the KK-variant magnetic field has been defined in Eq. \eqref{eq:magneticfield_KK-variant_def}.
This results exactly reproduces the transverse covariant 
electromagnetic current of the unbroken theory, as given in Eq.~\eqref{eq:final_constitutive_unbroken}. This contrasts with the Abelian model studied
in Ref.~\cite{Amado:2014mla}, in which the BZ term gives ${1\over 3}$ the result of the covariant current in the theory without spontaneous symmetry 
breaking.

\section{Closing remarks}
\label{sec:discussion}

In this paper we have studied the partition function and anomalous currents of a two-flavor chiral hadronic (super)fluid using the expressions
obtained in Ref. \cite{Manes:2018llx} in a general setup, particularized to the background of interest. In a first instance, we 
focused our attention onto the unbroken theory, computing the partition function 
from the anomalous functional realizing the Bardeen form of the anomaly. Explicit forms of the covariant currents were provided in terms of the
external sources, from where the transport coefficients
associated with the magnetic and chiral vortical effect are obtained. 
 The anomalous correction to the energy-momentum tensor of the unbroken theory
was also computed. 

In the case of the unbroken theory, we have explicitly computed the six transport coefficients in the constitutive relations
for the vector and axial-vector covariant gauge currents, as well as the anomaly-induced corrections to the energy-momentum tensor. 
When expressed in terms of KK-variant fields, we find no vorticity-dependent terms in the constitutive relations for the vector currents. 
This is the result of the specific structure of the U(2)$\times$U(2) group considered. 
Once written in terms of KK-invariant quantities, a nonvanishing chiral vortical conductivity appears, whose value is
determined by the geometry of the static background metric.

For the theory with chiral symmetry breaking, our computation of the covariant currents and the constitutive relations avoids the use
of the WZW effective action. Instead, we got the corresponding currents from the BZ terms of the unbroken theory, 
properly transformed by the matrix of NG bosons. Consistent currents were then easily evaluated by subtracting the contribution of
the BZ polynomials. Our method greatly simplifies the calculation of consistent (and covariant) currents, since expressions evaluated on the
particular background considered are used at every calculation step, instead of dealing with functional derivatives of the WZW action 
on a general background.

While the functional derivatives of the partition function lead to the consistent currents in equilibrium, out-of-equilibrium expressions of those 
currents are retrieved by considering their Lorentz covariant generalizations (see, e.g. \cite{Banerjee:2012iz}). 
In our case, we proceeded by writing the covariant currents in a Lorentz covariant form, to retrieve 
the desired constitutive relations for the hydrodynamics 
of a relativistic fluid. We have used this method to find the constitutive relations in the chiral hadronic fluid, 
with and without spontaneous symmetry breaking. 
In the former case, they are expressed in terms of five pseudo-scalars and four pseudo-vectors quantities depending on the 
NG fields, together
with the magnetic field and the vorticity vector. 
From them, we identify the pion field contributions to the chiral magnetic and vortical effects. Our calculation also predicts the emergence of a
chiral electric effect whose corresponding transport coefficient is explicitly evaluated. This effect contributes to all vector and axial-vector
covariant and consistent currents.

Whereas in the unbroken theory the chiral magnetic effect is restricted to the axial-vector currents, when expressed in terms of KK-variant
fields, we have found how
the presence of NG bosons induces this effect also in vector currents.  In all cases, its strength is 
controlled by the chemical potential governing chiral imbalance, and disappears in the limit $\mu_{5}\rightarrow 0$. A vorticity
dependent term survives nonetheless in the number densities, agreeing with similar contributions found in Ref.~\cite{Huang:2017pqe} using a 
different physical approach. As for the axial-vector global currents, they contain chiral separation effects associated with 
the electric and magnetic fields, as well as the vorticity. The later remains finite in the limit of vanishing chiral imbalance.

The strategy employed here to obtain the constitutive relations of a chiral (super)fluid can be extended to a wide class of theories with anomalous currents in both high energy and condensed matter physics. Gravitational and/or mixed gauge-gravitational anomalies~\cite{Jensen:2012kj}
can also be incorporated into the 
description by considering the appropriate anomaly polynomials. Due to the presence of curvature terms, they would contribute at
higher orders in the derivative expansion. Finding an economical and 
systematic way of computing the gravitational contributions to the constitutive relations
is indeed of interest, given their recently discovered experimental signatures~\cite{Gooth:2017mbd}. These and other issues will be addressed elsewhere.

\acknowledgments
We thank Claudio Corian\`o and Karl Landsteiner for discussions. The work of J.L.M. and M.V. has been supported by 
Spanish Science Ministry grant PGC2018-094626-B-C21 and Basque Government
grant IT979-16.  The research of E.M. is supported by Spanish MINEICO and European FEDER funds grants FIS2017-85053-C2-1-P, Junta de Andaluc\'{\i}a grant FQM-225, and Consejería de Conocimiento, Investigación y Universidad of
the Junta de Andalucía and European Regional Development Fund (ERDF)
(grant No. SOMM17/6105/UGR), as well as by Universidad del Pa\'{\i}s Vasco UPV/EHU through a Visiting Professor appointment and by Spanish MINEICO Ramón y Cajal Program (Grant No.
RYC-2016-20678). M.A.V.-M. acknowledges the financial support from the Spanish Science Ministry through research grant PGC2018-094626-B-C22, as well as from Basque Government
grant IT979-16. He also thanks the Department of Theoretical Physics of the University of the Basque Country 
for hospitality during the completion of this work.

\bibliographystyle{JHEP}
\bibliography{chiral_superfluid}

\end{document}